\documentclass[journal,letter,10pt,twocolumn]{IEEEtran}
\usepackage{amsmath,graphicx}

\usepackage{bbm}
\usepackage{amsmath, amssymb, amscd, amsthm, amsfonts}
\usepackage{subfiles}
\usepackage{color}

\usepackage[caption=false,font=footnotesize]{subfig}

\usepackage[backend=biber, style=ieee, dashed=false,isbn=false,doi=false,url = false,bibencoding=utf8]{biblatex}
\usepackage{hyperref}

\newtheorem{theorem}{Theorem}
\newtheorem{lemma}{Lemma}
\newtheorem{corollary}{Corollary}

\bstctlcite{IEEE:BSTcontrol}

\title{Modulo Sampling: Performance Guarantees in The Presence of Quantization}

\author{
   \IEEEauthorblockN{Neil Irwin Bernardo, \textit{Member, IEEE}, Shaik Basheeruddin Shah, \textit{Graduate Student Member, IEEE}, \\ and Yonina C. Eldar, \textit{Fellow, IEEE}
   }
    \thanks{Part of this work has been published in the conference proceedings of the IEEE International Symposium on Information Theory (ISIT) 2024.} 
    \thanks{N.I. Bernardo is with the Electrical and Electronics Engineering Institute, University of the Philippines Diliman, Quezon City 1101, Philippines (e-mail: neil.bernardo@eee.upd.edu.ph).}
    \thanks{S. B. Shah is with the Department of Electrical Engineering, Khalifa University, UAE (e-mail: shaik.shah@ku.ac.ae).}
    \thanks{Y. C. Eldar is with the Weizmann Institute of Science, Rehovot, Israel (e-mail: yonina.eldar@weizmann.ac.il).}
    \thanks{This research was supported by the Tom and Mary Beck Center for Renewable Energy as part of the Institute for Environmental Sustainability (IES) at the Weizmann Institute of Science, by the European Research Council (ERC) under the European Union’s Horizon 2020 research and innovation program (grant No. 101000967), by the European Union (ERC, SWIMS, 101119062), and by the Israel Science Foundation (grant No. 536/22). N.I. Bernardo acknowledges the Office of the Chancellor of the University of the Philippines Diliman, through the Office of the Vice Chancellor for Research and Development, for funding support through the PhD Incentive Award Grant 252509 YEAR 1. S. B. Shah acknowledges support from Khalifa University under the KU-Belgrade joint research collaboration.
   }
}
\begin{document}
%
\maketitle
\begin{abstract}
In this paper, we investigate the trade-off between dynamic range and quantization noise in modulo analog-to-digital converters (ADCs). We study two modulo ADC architectures: (1) a modulo ADC that outputs folded samples along with an additional 1-bit folding information signal, and (2) a modulo ADC that outputs only the folded samples. For the first system, we analyze a recovery algorithm that leverages the folding information to correctly unfold the quantized samples. Within the dithered quantization framework, we show that an oversampling factor $\mathrm{OF} > 3$ and quantizer resolution $b > 3$ are sufficient to guarantee the successful unfolding of the samples. Under these conditions, the system achieves a mean squared error (MSE) lower than that of a conventional ADC with the same number of amplitude quantization bits. Since folding information is typically unavailable in practical modulo ADCs, we also propose an orthogonal matching pursuit (OMP)-based recovery algorithm for the second system, which relies solely on the folded samples. In this case, we prove that the samples can be unfolded accurately if $\mathrm{OF} > 3$ and $b > 3 + \log_2(\zeta)$, for some penalty term $\zeta > 1$. For both systems, we show that the MSE of a modulo ADC is in $\mathcal{O}\left(\frac{1}{\mathrm{OF}^3}\right)$ when there is a sufficient number of bits for amplitude quantization. In contrast, the MSE of a conventional ADC is only in $\mathcal{O}\left(\frac{1}{\mathrm{OF}}\right)$. We further extend our analysis to the simultaneous acquisition of weak and strong signals occupying different frequency bands. Finally, numerical results are provided to validate the theoretical performance guarantees established.

\end{abstract}
\begin{IEEEkeywords}
Modulo ADC, sampling, quantization, weak and strong signals.
\end{IEEEkeywords}
\section{Introduction}
\label{section:intro}

Most modern data acquisition systems use analog-to-digital converters (ADCs) to transform continuous-time observations into a format suitable for digital processing. Standard ADCs first sample the signal at equally-spaced time intervals and then map the amplitude samples to a finite discrete set of values using a quantizer \cite{Eldar, Gray_1998}. The aim is to accurately represent the original analog signal in the digital domain. One critical consideration in an ADC design is ensuring that the dynamic range (DR) of the ADC exceeds that of the input signal, where DR refers to the difference between the smallest and largest values that can be represented. When the DR of the ADC falls short of capturing the full range of the input signal, clipping occurs \cite{HDR1, HDR2, HDR3}. This loss of information compromises the accuracy of the digital representation and impacts the reliability of subsequent data processing.
Moreover, obtaining a signal with lower DR can reduce the ADC's power consumption and can play a vital role in energy-constrained systems \cite{Bazzi:2025}.

\begin{figure}[t!]
    \centering
    \includegraphics[height = 5.5cm, width = 9.75cm]{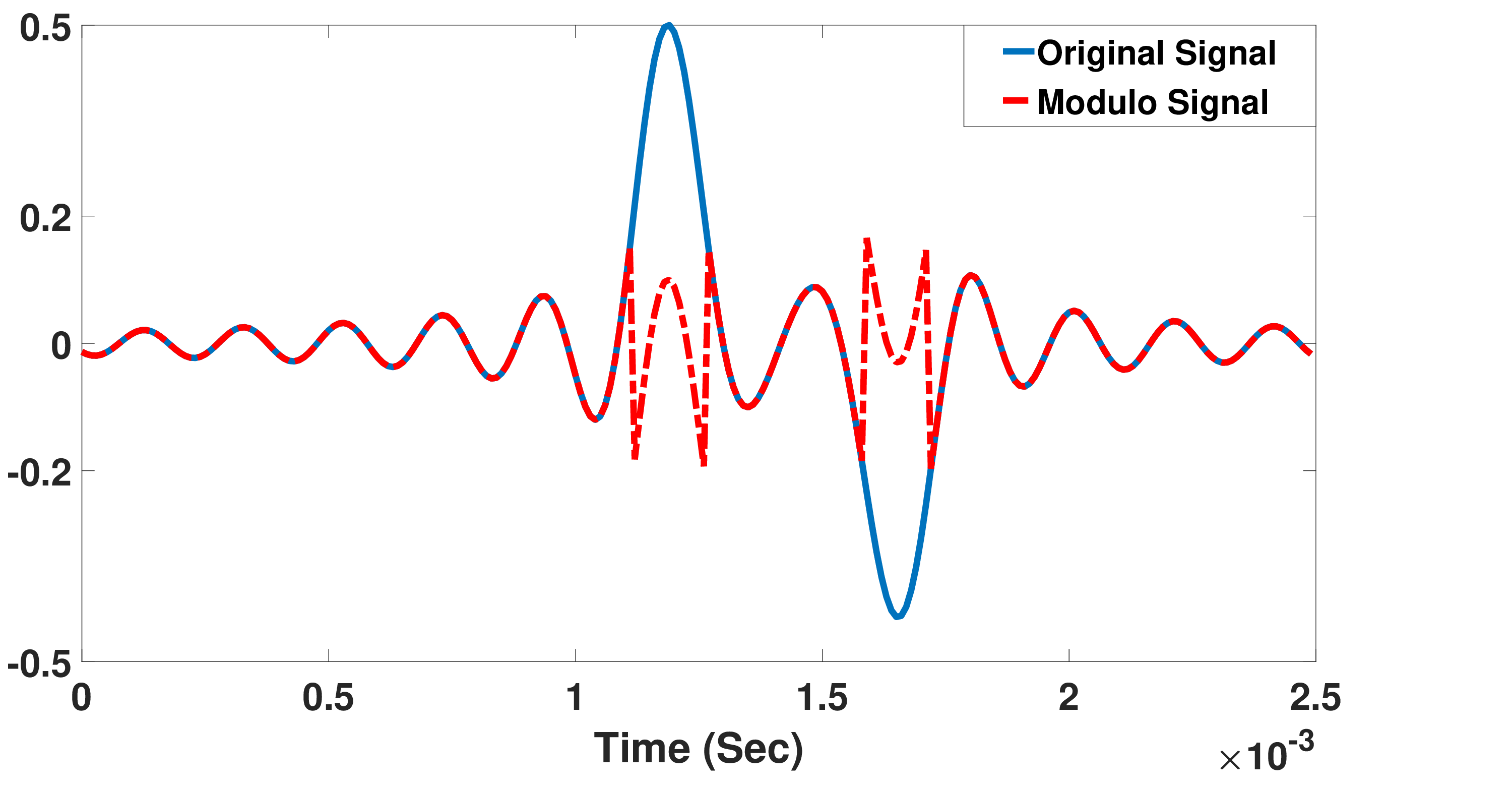}
    \caption{A typical high DR input signal and the corresponding modulo signal with modulo operator threshold value equal to $0.2$.}
    \label{Modulo_Example}
\end{figure}

There exist different algorithms in the literature to address this challenge \cite{Lit_1, Lit_4, Lit_3, Lit_2}.
The recent unlimited sampling framework (USF) has emerged as a promising solution \cite{Bhandari:2021} to enable the acquisition of high DR signals using limited DR ADCs. In USF, the high DR input signal is pre-processed by a non-linear modulo operator. This modulo operator folds the input signal whenever it crosses some pre-defined modulo thresholds (see Fig. \ref{Modulo_Example}). Subsequently, the modulo signal serves as the input to the ADC. Following the ADC, a recovery algorithm is applied to unfold the output samples of the ADC. The integration of an ADC, a modulo pre-processing, and a recovery algorithm is referred to as a modulo ADC system.

A key focus of existing research on modulo ADCs is developing robust, computationally efficient recovery algorithms that operate near the Nyquist rate. The existing literature presents diverse recovery algorithm strategies. These encompass higher-order-differences-based approach \cite{US1, Bhandari:2021}, prediction-based method \cite{Cheb}, wavelet-based technique \cite{Rudresh:2018}, Fourier-domain methodologies \cite{B2R2,Fourier-Prony,, B2R21, Romanov:2019,Bernardo2024SlidingDS}, compressed sensing-based approaches \cite{Shah:2023, Shah:2023b,Geethu:2025}, and others.
Among these methods, \cite{Shah:2023b,B2R2,B2R21,Shah:2023,Bernardo2024SlidingDS} are identified to be robust recovery algorithms approaching the Nyquist rate. Apart from recovery algorithms, there are also efforts to advance the hardware implementation of modulo ADCs \cite{Fourier-Prony, Mulleti:2023, Satish_HW, Shah:2023b, kvich2024practical, AyushHW,Zhu:2025}.
Furthermore, USF has been studied extensively across various signal models, such as Finite-Rate-of-Innovation (FRI) signals \cite{Mulleti:2024}, sparse signals \cite{Musa:2018, Prasanna:2021, Shah:2021}, graph signals \cite{Feng:2022}, multi-tone signals \cite{Zhang:2024}, and shift-invariant spaces \cite{Kvich:2024}, among others.

Nevertheless, research addressing the impact of quantization on modulo ADCs is notably limited. In \cite{US1}, the authors introduced a higher-order differences (HoD)-based recovery algorithm for bandlimited (BL) signals that operates at approximately $2\pi e\;(\approx 17)$ times the Nyquist rate when there is no quantization. Theoretical guarantees for the HoD-based recovery show that this sufficient condition for the sampling rate increases in the presence of bounded noise \cite{Bhandari:2021}. Recently, it was shown that this penalty in sampling rate under a bounded noise setting can be removed with high probability by using a multi-channel modulo sampler \cite{Florescu:2025}. To improve upon the HOD approach \cite{US1}, a prediction filter-based algorithm to unfold the modulo ADC output was proposed \cite{Cheb}. When the prediction filter is sufficiently long, this approach enables signal recovery at a rate close to the Nyquist rate. However, this recovery method fails in the presence of quantization noise.
In \cite{Florescu:2022}, the authors proposed a thresholding-based recovery algorithm with performance guarantees under bounded noise. However, their approach is based on a modulo ADC model that includes hysteresis and folding transients to account for hardware non-idealities. The recovery error guarantees established in \cite[Theorem 2]{Florescu:2022} become trivial under zero hysteresis. Recent work on USF radars \cite{Feuillen:2023} has investigated the impact of quantization on a modulo ADC if the input signal contains weak and strong components, however, this work uses HoD-based recovery to unfold the modulo samples. In \cite{krishna}, the authors focus on signal-to-quantization-noise ratio (SQNR) for inputs with varying amplitude distributions and emphasize hardware implementation, using a reconstruction method that relies on an encoded 2-bit reset signal.

This work aims to address this gap and provide a more comprehensive understanding of how modulo sampling and quantization interact, particularly in scenarios involving BL signals with diverse target strengths. Moreover, this study identifies suitable conditions for which a modulo ADC has better performance than a conventional ADC without modulo when the number of bits used for amplitude quantization is the same. The following are the major contributions of our work:

\begin{enumerate}
    \item We analyze the recovery algorithm \cite{Shah:2023b} that uses the quantized modulo samples and 1-bit folding information signal to unfold the modulo ADC output. In this setup, the 1-bit folding information\footnote{Note that the implementation in \cite{Shah:2023b} uses one bit from the $b$-bit ADC to detect modulo folding events. In contrast, this work uses a $b$-bit ADC for amplitude quantization and an additional 1-bit ADC to indicate folding events. The $b$ in \cite{Shah:2023b} needs to be increased by 1 to make the two modulo ADC implementations comparable in terms of bits per second.} indicates the positions of the modulo samples where folding occurred. Within a dithered quantization framework \cite{Schuchman:1964,Gray:1993, Gray_1998}, we evaluate the reconstruction performance of this approach. We establish sufficient conditions on the oversampling factor ($\mathrm{OF}$) and $b$ to ensure that the modulo ADC offers superior quantization noise suppression compared to conventional ADCs without the modulo operator. Specifically, we prove that $\mathrm{OF} > 3$ and $b > 3$ are sufficient for the modulo ADC to achieve lower mean squared error ($\mathrm{MSE}$) than conventional ADCs.

    \item Since the 1-bit folding information is generally unavailable in the modulo ADC, we propose and examine an orthogonal matching pursuit (OMP)-based recovery algorithm to unfold the modulo ADC output. Our analysis demonstrates that having $\mathrm{OF} > 3$ and $b > 3 + \log_2(\zeta)$, for some penalty term $\zeta > 1$, are sufficient for the modulo ADC without 1-bit folding information to achieve the same MSE performance guarantees as those of the modulo ADC with 1-bit folding information. Additionally, we present a case study that analyzes a BL signal having both weak and strong components occupying different frequency bands and demonstrate that a modulo ADC outperforms a conventional ADC in capturing the strong and weak signal components simultaneously.
    \item We show that, for the aforementioned settings, the MSE of the modulo ADC is on the order of $\mathcal{O}\left(\frac{1}{\mathrm{OF}^3}\right)$, while that of a standard ADC is on the order of $\mathcal{O}\left(\frac{1}{\mathrm{OF}}\right)$. This underscores the importance of oversampling in retrieving the original signal from quantized modulo observations. We validate the proposed theory with numerical simulations.
\end{enumerate}

A portion of the aforementioned contributions was previously presented in \cite{Bernardo_ISIT2024}, where a modulo ADC with explicit 1-bit folding information was analyzed within a dithered quantization framework. Building on this prior work, the present paper addresses the practical issue of modulo ADC without the 1-bit folding information by introducing an OMP-based signal recovery and establishing its MSE performance guarantees. Additionally,  we extend the analysis to BL signals with both strong and weak components.



The paper is organized as follows: The two modulo ADC system models considered in the study and our proposed recovery methods are described in Section II. Theoretical MSE performance guarantees and computational complexities of the recovery algorithms are established in Section III. Then, we specialize our derived performance guarantees to the case of simultaneous acquisition of weak and strong signals in Section IV. Numerical results are provided in Section V to validate our proposed theory. Finally, we summarize the work in Section VI.

\emph{Notation:} The following notations are used throughout the paper. We denote the set of real numbers, integers, and natural numbers as $\mathbb{R}$, $\mathbb{Z}$, and $\mathbb{N}$, respectively.
The space of square-integrable functions on $\mathbb{R}$ is represented by $L^2(\mathbb{R})$, and $\mathbb{E}[\cdot]$ denotes the expectation operator.
We use bold lowercase letters to represent vectors, and bold capital letters to denote matrices. The $\ell_p$-norm of a vector $\mathbf{z}$ is denoted as $\|\mathbf{z}\|_{p}$, with a similar notation extending to $\ell_{p}$-norms of functions. When referring to discrete-time signals, we employ the notation $z[n]$ to signify $z(nT_{\mathrm{s}})$, assuming the sampling period $T_{\mathrm{s}}$ is evident from the context. For a discrete-time signal $x[n]$, the first-order difference operator is defined as $\Delta x[n] = x[n] - x[n-1]$, under the assumption that $x[-1] = 0$.
To describe the growth rate of a quantity concerning some parameter, we employ standard asymptotic notation, such as $\mathcal{O}(\cdot)$. The cumulative sum operation applied to a vector $\mathbf{x}\in\mathbb{R}^{N\times 1}$ yields a vector $\mathbf{y}\in\mathbb{R}^{N\times 1}$, where the $k^{\text{th}}$ element of $\mathbf{y}$ is given by $y_k = \sum_{i = 1}^k x_i$.
The pseudo-inverse of a matrix $\mathbf{A} \in \mathbb{R}^{m \times n}$, where $n\gg m$, is denoted by $\mathbf{A}^{\dagger}$, and defined as $\mathbf{A}^{\dagger} = (\mathbf{A}^T\mathbf{A})^{-1}\mathbf{A}^T$. We used $\omega$ and $\Omega$ to denote the frequencies for the CTFT and DTFT spectrum, respectively.

\section{Modulo ADC Problem Setup and Proposed Recovery Methods}
\label{section:system-model}

In this section, we describe the two modulo ADC system models to be analyzed in this work: (1) a modulo ADC that generates the quantized folded samples and an extra 1-bit folding information, and (2) a modulo ADC without the extra 1-bit folding information. We present the proposed recovery algorithms for these two system models.

\subsection{Modulo ADC System With Extra 1-Bit Folding Information}
\label{subsection:modulo_ADC_A}

\begin{figure*}[t!]
    \centering
    \subfloat[]{
    \includegraphics[scale = .975]{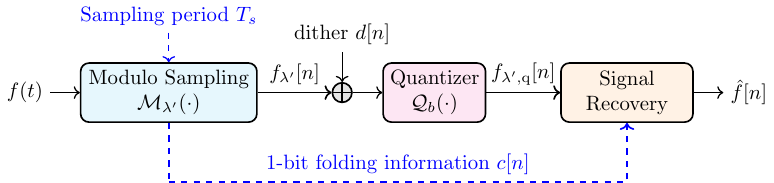}
    \label{fig:modulo_ADC_a}
    }

    \subfloat[]{
    \includegraphics[scale = .975]{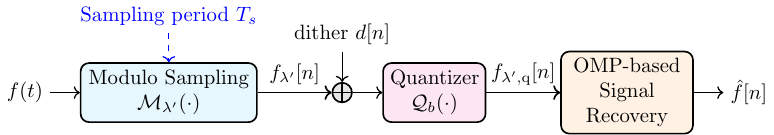}
    \label{fig:modulo_ADC_b}
    }
    \caption{Schematic diagram of (a) a modulo sampling system with dithered quantization. A 1-bit folding information signal is also generated by $\mathcal{M}_{\lambda'}(\cdot)$ to aid in signal reconstruction. In (b), the extra 1-bit folding information signal is removed.}
\end{figure*}
We consider the modulo ADC system shown in Fig. \ref{fig:modulo_ADC_a}. The input $f(t)\in L^1(\mathbb{R})$ is a bandlimited signal with frequency support $\left(-\frac{\omega_m}{2},\;+\frac{\omega_m}{2}\right)$. The input signal is fed to a non-linear modulo sampling block, denoted $\mathcal{M}_{\lambda'}(\cdot)$, to produce the discrete-time modulo observations $f_{\lambda'}[n],\;n\in\mathbb{Z}$. Mathematically, 
\begin{align}\label{eq:modulo_operator}
    f_{\lambda'}[n] &=\mathcal{M}_{\lambda'}\left(f(t)\right) \nonumber\\
    &= [(f(nT_s)+\lambda')\;\mathrm{mod}\;2\lambda'] - \lambda',
\end{align}
where $\lambda'\in(0,\|f(t)\|_{\infty})$ is the modulo operator threshold. The range of $\lambda'$ is chosen to exclude the trivial case that the input signal amplitude is within the linear region of the modulo. The sampling rate is $\omega_s = \frac{2\pi}{T_s} = \mathrm{OF}\times \omega_m$, where $\mathrm{OF} > 1$. We define $\rho = \frac{1}{\mathrm{OF}}$ for our analysis. Effectively, the frequency support of $f[n] = f(nT_{\mathrm{s})}$ is $(-\rho\pi,+\rho\pi)$.


The folded samples are fed to a $b$-bit scalar quantizer $\mathcal{Q}_b(\cdot)$ with dynamic range $[-\lambda,+\lambda]$. The nonlinearity of the quantization process makes it difficult to analyze the impact of quantization in modulo sampling. To this end, a dither signal $d[n]$ is added to $f_{\lambda'}[n]$ prior to quantization. An independent and identically-distributed (i.i.d.) triangle noise sequence with amplitude support $\big(-\frac{2\lambda}{2^b},+\frac{2\lambda}{2^b}\big]$ is used for the dither signal $d[n]$. The rationale for incorporating a triangle dither in the system model is twofold. First, the modulo operation can be set so that $f_{\lambda'}[n] + d[n]$ will not overload the quantizer. This, together with the properties of triangle dither, guarantees that Shuchman conditions \cite{Schuchman:1964} are satisfied, which allows the first-order and second-order statistics of the quantization noise to be derived from the system parameters. Second, using \cite[Theorem 2]{Gray:1993}, the sequences $\epsilon[n]$ and $f_{\lambda'}[n]$ become uncorrelated and the conditional second-order moment of $\epsilon[n]$ becomes independent of $f_{\lambda'}[n]$, i.e., $\mathbb{E}\left[\epsilon^2[n]|f_{\lambda'}[n]\right] = \mathbb{E}\left[\epsilon^2[n]\right]$.


Each quantization bin has a width of $\frac{2\lambda}{2^b}$. To prevent the quantizer from being overloaded by the folded signal, we set the ADC DR to be $\lambda = \frac{2^b \lambda'}{(2^b-2)}$. The quantizer output can then be written as
\begin{align}
        f_{\lambda',\mathrm{q}}[n] =& \mathcal{Q}_{b}\left(f_{\lambda'}[n] + d[n]\right) \nonumber \\
        =& f[n] + z[n] + \epsilon[n],
\end{align}
where $z[n] \in 2\lambda'\mathbb{Z}$ is the residual samples due to the folding operation of $\mathcal{M}_{\lambda'}(\cdot)$ and $\epsilon[n] = \mathcal{Q}_{b}\left(f_{\lambda'}[n] + d[n]\right) - f_{\lambda'}[n]$ is the quantization noise sequence. With triangle dither, the quantization noise is a white process with noise power \cite{Gray:1993}
\begin{equation}
    \mathbb{E}\left[\epsilon^2[n]\right] = \frac{1}{4}\left(\frac{2\lambda}{2^b}\right)^2 = \frac{\lambda^2}{2^{2b}}.
\end{equation} 
We will use the statistical properties of $\epsilon[n]$ to establish MSE performance guarantees in Section III. However, note that the inclusion of a triangular dither signal is mainly for analytical tractability. 
We show in Section \ref{section:numerical} that our proposed recovery methods work even without the dither signal.

In addition to $f_{\lambda'}[n]$, the modulo sampling block also generates a 1-bit discrete-time binary signal $c[n]$ which contains information about the $(2\mathbb{Z} + 1)\lambda'$ level crossings in $f(t)$. More precisely, $c[n] = 1$ whenever the input signal crosses the level $(2\mathbb{Z}+1)\lambda'$ within the time interval $(nT_s,(n+1)T_s]$ while $c[n] = 0$ otherwise. The set of indices $n$ where $c[n] = 1$ is denoted as $\mathcal{S}$. 

There are several ways to generate this 1-bit folding information. One approach is explored in \cite{Shah:2023b}, which only incurs one additional OR gate to the setup. This penalty is fixed regardless of $b$. From a power efficiency viewpoint, this additional OR gate has minimal impact on the power consumption of the modulo ADC. In FPGA implementations, the OR gate can be realized using a small portion of a look-up table (LUT), resulting in negligible resource usage and power overhead. In ASIC implementations, it can be built using standard CMOS logic, typically requiring only a few transistors with minimal area and energy consumption. Since these implementation paths are straightforward and well-supported by existing digital design flows, the prototype in \cite{Shah:2023b} remains practical and scalable, even when extended to higher-bit resolutions. However, we note that challenges may arise in very high-speed applications where generating this 1-bit side information and synchronizing it with the folded samples become more difficult.

\subsection{Signal Recovery for Modulo ADC With 1-Bit Folding Information}
\label{subsection:recovery_method_A}

The signal recovery block utilizes the information about $\mathcal{S}$ to recover the modulo residue $z[n]$ in the frequency domain. Let $\Delta f_{\lambda',\mathrm{q}}[n] = \Delta f[n] + \Delta z[n] + \Delta \epsilon[n]$ be the first-order difference of $f_{\lambda',\mathrm{q}}[n]$. The discrete-time Fourier Transform (DTFT) of $\Delta f_{\lambda',\mathrm{q}}[n]$ for $\Omega\in(\rho\pi,2\pi-\rho\pi)$ can be expressed as
\begin{align}\label{eq:F_OOB_DTFT}
    F_{\Delta,\mathrm{OOB}}(e^{j\Omega}) &= \sum_{n = -\infty}^{\infty}\Delta f_{\lambda',\mathrm{q}}[n]e^{-j\Omega n}\nonumber\\
    &= \sum_{n = -\infty}^{\infty} \left[\Delta z[n] + \Delta\epsilon[n]\right]e^{-j\Omega n}.
\end{align}
The subscript $\mathrm{OOB}$ is used to denote that we are only interested in out-of-band (OOB) frequencies $\Omega\in(\rho\pi,2\pi-\rho\pi)$. The second line follows since the signal $f[n]$ is within $(-\rho\pi,\; +\rho\pi)$. As such, $\Delta f[n]$ has no contribution to $F_{\Delta,\mathrm{OOB}}(e^{j\Omega})$. Because $f(t) \in L^1(\mathbb{R})$, we have $\lim_{|t|\rightarrow\infty}f(t) = 0$ by the Reimann-Lebesgue lemma \cite[Chapter 12]{gradshteyn2007}. Consequently, for any $\lambda' > 0$, there exist integers $n_0$ and $n_1$ ($n_0 < n_1$) such that $|f(nT_{\mathrm{s}})|< \lambda$ for all $n_0 > n$ and $n_1 \leq n$. The first-order difference of the residual samples, $\Delta z[n]$, can be treated as a finite-duration discrete-time signal of length $N = n_1 - n_0$. Equation \eqref{eq:F_OOB_DTFT} can be written as
\begin{align}\label{eq:F_OOB_DTFT_2}
    F_{\Delta,\mathrm{OOB}}(e^{j\Omega}) =& \sum_{n = n_0}^{n_1-1} \left[\Delta z[n] + \Delta\epsilon[n]\right]e^{-j\Omega n}\nonumber \\
    &+ \sum_{n \notin \{n_0,\cdots,n_1-1\}} \Delta\epsilon[n]e^{-j\Omega n}.
\end{align}
We approximate $F_{\Delta,\mathrm{OOB}}(e^{j\Omega})$ by setting $f[n] = 0$ for $n < n_0$ and $n_1$ and defining $y_{k}$ to be
\begin{align}
    y_{k} =& \frac{1}{\sqrt{N}}\sum_{n = n_0}^{n_1-1} \left[\Delta z[n] + \Delta\epsilon[n]\right]e^{-j \frac{2\pi k (n-n_0)}{N}}\nonumber\\
    =&\frac{1}{\sqrt{N}}\Big(\sum_{n=n_0,n \in \mathcal{S}}^{n_1-1}\left[\Delta z[n]+ \Delta\epsilon[n]\right]e^{-j\frac{2\pi k(n-n_0)}{N}}\nonumber\\
             &\quad+ \sum_{n = n_0,n\notin \mathcal{S}}^{n_1-1}\Delta\epsilon[n]e^{-j\frac{2\pi k (n-n_0)}{N}}\Big).
\end{align}

Here, $\frac{2\pi k}{N}$ is one of the $K < N$ discrete frequencies inside $(\rho\pi,\; 2\pi - \rho\pi)$. Effectively, $y_{k}$ is the normalized DFT of the truncated $\Delta f_{\lambda',\mathrm{q}}[n]$ in the OOB region. The second line follows from putting all terms with $n\in\mathcal{S}$ in the first summation and putting the remainder in the second summation. Notice that the residual samples only appear in the first summation. We vectorize $\Delta z[n]$ by forming the vector ${\Delta}\mathbf{z} = [\Delta z[n_0],\cdots,\;\Delta z[n_1-1]]^T$. We also define the $|\mathcal{S}|\times 1$ vector ${\Delta}\mathbf{z}_{\mathcal{S}}$ which contains the nonzero elements of ${\Delta}\mathbf{z}$ at indices found in the set $\mathcal{S}$ while the $(N-|\mathcal{S}|)\times 1$ vector ${\Delta}\mathbf{z}_{\mathcal{S}^{c}}$ is an all-zero vector.

Let $\mathbf{V}\in \mathbb{R}^{K\times N}$ with elements $\mathbf{V}_{k,n} = e^{-j\frac{2\pi k (n-n_0)}{N}}$, where $\frac{2\pi k}{N}$ is the $k$-th discrete frequency in $(\rho\pi, 2\pi - \rho\pi)$. The matrix $\mathbf{V}_{\mathcal{S}}\in \mathbb{R}^{K\times |\mathcal{S}|}$ contains the $|\mathcal{S}|$ columns of $\mathbf{V}$ corresponding to the indices in $\mathcal{S}$. An estimate of the non-zero elements of the first-order difference vector, denoted ${\Delta}\mathbf{\hat{z}}_{\mathcal{S}}$, can be formed by
\begin{align}
            {\Delta}\mathbf{\hat{z}}_{\mathcal{S}} = \mathbf{V}_{\mathcal{S}}^{\dagger}\mathbf{y}.
\end{align}
The estimate of the vector ${\Delta}\mathbf{z}$, denoted ${\Delta}\mathbf{\hat{z}}$, is obtained by combining ${\Delta}\mathbf{\hat{z}}_{\mathcal{S}}$ and ${\Delta}\mathbf{\hat{z}}_{\mathcal{S}^{c}}$. The elements of ${\Delta}\mathbf{\hat{z}}$ are rounded to the nearest $2\lambda'\mathbb{Z}$. Then, a cumulative sum is applied to the resulting vector to obtain the estimate of the modulo residue, denoted $\mathbf{\hat{z}} = ({\Delta}\mathbf{\hat{z}})_{\Sigma}$. Let $\hat{z}[n]$ be the discrete-time representation of the vector $\mathbf{\hat{z}}$. We subtract $\hat{z}[n]$ from $f_{\lambda',\mathrm{q}}[n]$ to remove the modulo residue from the quantized modulo observations. Finally, we apply an ideal digital lowpass filter, $\mathrm{LPF}\{\cdot\}$, with passband region $\left(-\frac{\omega_m}{2},\;+\frac{\omega_m}{2}\right)$. The recovered signal can be written as
\begin{align}\label{eq:recovered_sig}
    \hat{f}[n] =& \mathrm{LPF}\left\{f_{\lambda',\mathrm{q}}[n] - \hat{z}[n]\right\}\nonumber\\ 
    =& f[n] + \mathrm{LPF}\left\{z[n] - \hat{z}[n]\right\} + \mathrm{LPF}\{\epsilon[n]\},
\end{align}
for $n \in \{n_0,\cdots,n_1-1\}$. Hence, the recovered signal is composed of the desired samples, filtered modulo residue estimation error, and filtered quantization noise.

\subsection{Modulo ADC Without 1-Bit Folding Information}
\label{subsection:recovery_method_B}

In a typical modulo ADC system, the 1-bit folding information signal may not be available for the reconstruction method. Thus, it is also of interest to develop a recovery algorithm for the modulo ADC shown in Fig. \ref{fig:modulo_ADC_b}. Since the 1-bit folding information tells the exact locations of nonzero elements of $\Delta z$, one approach to facilitate the signal reconstruction is to perform support recovery of $\Delta z[n]$ and then run the signal reconstruction algorithm described in Section \ref{subsection:recovery_method_A}, with $c[n]$ being replaced by $\hat{c}[n]$, the recovered support of $\Delta z[n]$. Note that $\Delta z[n]$ is a sparse signal \cite[Section III]{Shah:2023}. Thus, we can utilize existing support recovery algorithms for sparse signals and their corresponding theoretical guarantees.

For the support recovery of $\Delta z[n]$, we consider the greedy \emph{orthogonal matching pursuit} (OMP) algorithm \cite{Mallat:1993, Tropp_2007, LIP2}, which uses a stopping rule based on the $\ell_{\infty}$-norm of the bounded noise \cite{Cai:2011}. More precisely, $\hat{c}[n]$ is obtained as follows:
\begin{enumerate}
    \item Initialize the index set $\mathcal{I} = \emptyset$, the residue $\mathbf{r}_0 = \mathbf{y}$, the estimator $\mathbf{x}_0 = \mathbf{0}$, and the iteration counter $t = 1$.
    \item At time $t$, select a column from $\mathbf{V}$ that is most correlated with the current residue $\mathbf{r}_{t-1}$, i.e.,
    \[n_{t} = \underset{n}{\arg\max} |\left<\mathbf{v}_n,\mathbf{r}_{t-1}\right>|,\]
    where $\mathbf{v}_{n}$ is the $n$-th column of $\mathbf{V}$. The index $n_{t}$ is added to the index set $\mathcal{I}$.
    \item Calculate the new estimator $\mathbf{x}_t$ by projecting $\mathbf{y}$ onto the space spanned by $\mathbf{V}_{\mathcal{I}}$, i.e., $\mathbf{x}_t = \mathbf{V}_{\mathcal{I}}^{\dagger}\mathbf{y}$. Update residual $\mathbf{r}_t = \mathbf{y} - \mathbf{V}\mathbf{x}_t$
    \item Stop if $\|\mathbf{V}^{\mathrm{H}}\mathbf{r}_{t}\|_{\infty}\leq \eta\frac{6\lambda'}{2^{b}-2}$, where $\eta = \|\mathbf{V}^{\mathrm{H}}\mathbf{V}\|_{\infty}$ and set
    \begin{align*}
        \hat{c}[n] = \begin{cases}
            1,\;n\in\mathcal{I}\\
            0,\;n\notin\mathcal{I}
        \end{cases}.
    \end{align*}
    Otherwise, set the iteration counter to $t = t+1$ and return to Step 2.
\end{enumerate}
The stopping criterion in step 4 is used so that the residue is less than the $\ell_{\infty}$-norm of $\Delta \boldsymbol{\epsilon}$. Once $c[n]$ has been estimated, we can apply the recovery algorithm in Section \ref{subsection:recovery_method_A} accordingly.

Our aim is to derive MSE performance guarantees for the two modulo ADC systems presented when the proposed recovery algorithms are used. The following section establishes these performance guarantees.




\section{Recovery Guarantees and Computational Complexity}
\label{section:performance_guarantee}


We quantify the recovery performance of the modulo ADC using the MSE criterion, which can be mathematically expressed as
\begin{align}
    \mathrm{MSE} =& \frac{1}{N}\sum_{n = n_0}^{n_1-1}\mathbb{E}\left\{\big|\hat{f}[n] - f[n]\big|^2\right\}\nonumber\\
    &+ \sum_{n\notin\{n_0,\cdots,n_1-1\}}\mathbb{E}\{|f[n]|^2\},
\end{align}
where $N = n_1 - n_0$, as defined earlier. The second summation accounts for the truncation error, denoted $\mathrm{MSE}_{t}$, due to neglecting the signal values outside $\{n_0,\cdots,n_1-1\}$ while the first summation accounts for the reconstruction error, denoted $\mathrm{MSE}_{r}$, due to the modulo residue estimation error and finite quantization. In this paper, we focus on the latter since we can set $n_0$ and $n_1$ so that $f[n] \approx 0$ for $n\notin\{n_0,\cdots,n_1-1\}$. Consequently, the truncation error can be set to a negligible value. Henceforth, we refer to the reconstruction error whenever MSE is mentioned. 
The reconstruction error can be expressed as
\begingroup
\allowdisplaybreaks
\begin{align}\label{eq:reconstruction_error}
    \mathrm{MSE}_r =& \frac{1}{N}\sum_{n = n_0}^{n_1-1}\mathbb{E}\left\{\big|\hat{f}[n] - f[n]\big|^2\right\}\nonumber\\
    =& \frac{1}{N}\sum_{n = n_0}^{n_1-1}\bigg|\mathrm{LPF}\left\{z[n] - \hat{z}[n]\right\}\bigg|^2\nonumber\\
    &+2\mathbb{E}\left\{\mathrm{LPF}\left\{\epsilon[n]\right\}\right\}\mathrm{LPF}\left\{z[n] - \hat{z}[n]\right\}\nonumber\\
    &+\mathbb{E}\left\{\big|\mathrm{LPF}\left\{\epsilon[n]\right\}\big|^2\right\}\nonumber\\
    =& \frac{1}{N}\sum_{n = n_0}^{n_1-1}\bigg|\mathrm{LPF}\left\{z[n] - \hat{z}[n]\right\}\bigg|^2+\mathbb{E}\left\{\big|\mathrm{LPF}\left\{\epsilon[n]\right\}\big|^2\right\}.
\end{align}
\endgroup
The second line follows from expanding the expression in the first line. Since $\epsilon[n]$ has zero mean, the second term of the second line can be dropped. The first term of the third line comes from the lowpass-filtered modulo residue estimation error while the second term of the third line comes from the in-band quantization noise. Our objective in this section is to derive sufficient conditions for $\mathrm{OF}$ and $b$ so that in-band quantization noise is the only source of reconstruction error for the modulo ADC with 1-bit folding information and modulo ADC without 1-bit folding information.

\subsection{MSE guarantee for Modulo ADC With 1-Bit Folding Information}

We first focus on the MSE performance guarantee for the modulo ADC with 1-bit folding information. One important property to ensure correct signal unfolding is to show that $\mathbf{V}_{\mathcal{S}_i}$ is a full column rank matrix. With $\mathbf{V}_{\mathrm{S}}$ being full column rank, there is a unique solution to the equation $\mathbf{y} = \mathbf{V}_{\mathcal{S}}\cdot{\Delta}\mathbf{z_{\mathcal{S}}}$. The following key lemma shows that an OF value greater than three is a sufficient (but not necessary) condition to make $\mathbf{V}_{\mathcal{S}}$ a full column rank matrix. 
\begin{lemma}\label{lemma:oversampling_req}
Suppose $\mathrm{OF} > 3$ and $n_0$,$n_1$ are chosen so that truncation error is negligible. Then $\mathbf{V}_{\mathcal{S}}$ has full column rank if the modulo threshold $\lambda' \geq \frac{\|f(t)\|_{\infty}}{\mathrm{OF}-2}$.

\end{lemma}
\begin{proof}
    See Appendix \ref{proof:appendix_A}.
\end{proof}
The intuition of the proof is to relate the oversampling factor $\mathrm{OF}$ to the dimensions of the sensing matrix $\mathbf{V}_{\mathcal{S}}$. Specifically, increasing $\mathrm{OF}$ has two key effects:
(1) it increases the number of discrete frequencies in the out-of-band (OOB) region, thereby increasing the number of rows in $\mathbf{V}_{\mathcal{S}}$; and
(2) it reduces the number of folding events in a length-$N$ signal, which corresponds to a decrease in the number of columns of $\mathbf{V}_{\mathcal{S}}$.
The proof identifies a regime where the number of rows $K$ exceeds or matches the number of columns $|\mathcal{S}|$, ensuring that the sensing matrix is sufficiently tall to enable stable recovery under appropriate conditions.

For a given $f(t)$, it actually suffices that $\mathrm{OF} \geq \frac{N}{N - |\mathcal{S}|}$ to make $\mathbf{V}_{\mathcal{S}}$ full column rank. However, such a condition depends on the specific $f(t)$. By using the upper bound on $|\mathcal{S}|$ established in \cite{Shah:2023}, we eliminate the dependence on the specific $f(t)$ and find an oversampling factor requirement that holds for any $f(t)$. However, the bound on $|\mathcal{S}|$ is not tight in general as demonstrated in \cite[Table 1]{Shah:2023}. 

We are now ready to state the main result of this paper. The following theorem gives the exact MSE of the modulo ADC system in Fig. \ref{fig:modulo_ADC_a} when $b > 3$, $\mathrm{OF} > 3$, and $\lambda'$ is set to be $\frac{\|f(t)\|_{\infty}}{\mathrm{OF}-2}$.

\begin{theorem}\label{theorem:quant_noise}
    Suppose $b > 3$, $\mathrm{OF} > 3$, and $n_0$,$n_1$ are chosen so that truncation error is negligible. Then the $\mathrm{MSE}_{r}$ of the reconstruction procedure for the modulo ADC with extra 1-bit folding information is
    \begin{align}\label{eq:MSE_r_1}
        \mathrm{MSE}_{r} = \frac{c^2}{\mathrm{OF}(2^b - 2)^2}\cdot\left(\frac{\|f(t)\|_{\infty}}{\mathrm{OF}-2}\right)^2
    \end{align}
    if we set $\lambda' = \frac{c\cdot\|f(t)\|_{\infty}}{\mathrm{OF}-2}$ for some positive constant $c \in [1,\mathrm{OF} - 2]$.
\end{theorem}
\begin{proof}
    See Appendix \ref{proof:appendix_B}.
\end{proof}

The intuition of the proof is to set a sufficiently large number of quantization bits such that the induced quantization noise is not strong enough to put the recovered first-order difference of the residue outside $(2\lambda'p-\lambda',2\lambda'p + \lambda')$ if $z[n] = 2\lambda'p$ for some $p\in\mathbb{Z}$. In this case, the rounding operation on the elements of ${\Delta}\hat{z}$ is perfect, i.e., $\hat{z}[n] = z[n]$, and the only impairment in $\hat{f}[n]$ is the (lowpass-filtered) quantization noise.

We now compare our derived MSE for the modulo ADC in Fig.~\ref{fig:modulo_ADC_a} with that of the conventional ADC without modulo sampling. To get the behavioral model of a conventional ADC, we simply replace $\mathcal{M}_{\lambda'}(\cdot)$ in Fig. \ref{fig:modulo_ADC_a} with a regular sampler with sampling period $T_{\mathrm{s}}$ and then remove $c[n]$. To obtain analytical results for conventional ADCs, we also use the non-subtractive dithered quantization framework.  With triangle dither $d[n]\in \left(-\frac{2\lambda}{2^b},\;+\frac{2\lambda}{2^b}\right]$, we must set $\lambda$ of the conventional ADC as $\lambda = \frac{2^b}{2^b - 2}\|f(t)\|_{\infty}$ to prevent overloading. Consequently, the quantization noise is white and has a mean squared value of $\mathbb{E}\left[\epsilon^2[n]\right] = \frac{\lambda^2}{2^{2b}}$. With an oversampling factor of $\mathrm{OF}$, the desired signal occupies only $\frac{1}{\mathrm{OF}}$ of the whole DTFT spectrum. A lowpass filter with passband region $(-\rho\pi,+\rho\pi)$ can be used to filter out the quantization noise in the out-of-band region. Thus, the MSE is solely attributed to the filtered quantization noise and can be expressed as
\begin{align}\label{eq:no_modulo_error}
    \mathrm{MSE}^{(\mathrm{no-modulo})} = \frac{1}{\mathrm{OF}(2^b-2)^2}\cdot \|f(t)\|_{\infty}^{2}.
\end{align}
By comparing \eqref{eq:no_modulo_error} with the result in Theorem \ref{theorem:quant_noise}, we see that, for the same number of bits for amplitude quantization, the quantization noise power of the modulo ADC is strictly lower than that of the conventional ADC when $\mathrm{OF} > 3$ and $b > 3$. In fact, it can be observed that $\mathrm{MSE}^{(\mathrm{no-modulo})} = \mathcal{O}\left(\frac{1}{\mathrm{OF}}\right)$ whereas $\mathrm{MSE}^{(\mathrm{with-modulo})} = \mathcal{O}\left(\frac{1}{\mathrm{OF}^3}\right)$. This sheds light on the advantage of modulo ADCs over conventional ADCs.

We also compare our results with other theoretical guarantees derived for modulo sampling. The impact of bounded noise (e.g., quantization) has been analyzed in \cite[Theorem 3]{Bhandari:2021} when the recovery algorithm is based on HoD. Their analysis guarantees the recovery of the samples up to an unknown additive constant, i.e., $\hat{f}^{\mathrm{(HOD)}}[n] = f[n] + \epsilon[n] + 2\lambda'p$ for some unknown $p\in\mathbb{Z}$. By ignoring the unknown additive constant and applying an appropriate noise filtering, the MSE of the HoD should be identical to that of our proposed recovery algorithm. However, under this bounded noise setting, the sufficient condition to achieve this performance guarantee is that the sampling rate should be at least $2^{\alpha}\pi e$ times the Nyquist rate, where $\alpha\in\mathbb{N}$ is a noise-dependent parameter. In contrast, our analytical result does not have an unknown additive constant and the exact MSE can be derived at any $\mathrm{OF} > 3$.  A prediction-based method introduced in \cite{Romanov:2019} demonstrated that the oversampling factor (\(\mathrm{OF}\)) can be reduced arbitrarily close to 1 by significantly increasing the length of the prediction filter. However, as noted in \cite[Section~III]{Romanov:2019}, the method becomes unreliable when quantization noise is present. In contrast, our approach remains effective even when using a finite number of quantization bits. 

We would also like to mention key differences between our proposed approach and the UNO framework \cite{UNO}. Unlike UNO, which relies on a large number of 1-bit measurements per sample using varying thresholds, our method processes a single quantized measurement per folded sample. Furthermore, our reconstruction guarantees are deterministic and hold for all finite-energy bandlimited signals, while UNO provides high-probability bounds. Lastly, our oversampling requirement is significantly lower, with a sufficient condition of $\mathrm{OF} \geq 3$ compared to $\mathrm{OF} \geq 2^{h}\pi e$, where $h\in\mathbb{N}$, required in \cite{UNO}. We also observe that the condition $\mathrm{OF} > 3$ is similar to the oversampling requirement established in \cite{Zhang:2024b} for the unique recovery of periodic bandlimited signals under a modulo-DFT sensing framework. Unlike our model, their approach applies the DFT to the sampled signal before the modulo operation—a structure that may be impractical to implement in real-world systems. Finally, we also note that sliding DFT approach of our proposed algorithm has been introduced in \cite{Bernardo2024SlidingDS}. The recovery guarantees established in \cite{Bernardo2024SlidingDS} adopted the recovery guarantees in the conference version of our work \cite{Bernardo_ISIT2024} but incorporated the impact of spectral leakage in the algorithm performance. In contrast, we assumed that spectral leakage can be neglected in our work if $n_0$ and $n_1$ are chosen appropriately.



\subsection{MSE Guarantee for Modulo ADC Without 1-Bit Folding Information}

We now consider the modulo ADC system in Fig. \ref{fig:modulo_ADC_b} and establish a recovery performance guarantee for the proposed algorithm in Section \ref{subsection:recovery_method_B}. Since the proposed algorithm is based on OMP, we restate a theorem from \cite{Wu:2013} which identifies a sufficient condition on the minimum amplitude of the noisy sparse signal to ensure perfect support recovery.

\begin{theorem}{(from \cite[Theorem 2]{Wu:2013})} \label{theorem:amplitude_req} Consider the observation signal $\mathbf{y} = \mathbf{A}\mathbf{x} + \mathbf{e}$. Suppose that $\|\mathbf{A}^{\mathrm{H}}\mathbf{e}\|_{\infty}\leq \epsilon_0$ and matrix $\mathbf{A}$ satisfies
\begin{align}
    \delta_{L+1} < \frac{1}{\sqrt{L}+1},
\end{align}
where $\delta_{L}$ is the \emph{restricted isometry property} (RIP) constant order $L$ of matrix $\mathbf{A}$. Then, the OMP with stopping rule $\|\mathbf{A}^{\mathrm{H}}\mathbf{e}\|_{\infty} \leq \epsilon_0$ will exactly recover the support $\Omega$ of an $L$-sparse signal $\mathbf{x}$ from the observation signal $\mathbf{y}$ if the minimum magnitude of nonzero elements of $\mathbf{x}$ satisfies
\begin{align}\label{eq:OMP_req}
    \min_{i\in\Omega}|\mathbf{x}_{i}| > \frac{\left(\sqrt{1+\delta_{L+1}}+1\right)\sqrt{L}\cdot\epsilon_0}{1-(\sqrt{L}+1)\delta_{L+1}}.
\end{align}
\end{theorem}

Using Theorem \ref{theorem:amplitude_req}, we identify a sufficient condition on the quantizer resolution $b$ to ensure that OMP will produce $\hat{c}[n] = c[n]$. With perfect $\hat{c}[n]$, we can apply Theorem \ref{theorem:quant_noise} to obtain recovery guarantees for the proposed algorithm in Section \ref{subsection:recovery_method_B}.

\begin{theorem}\label{theorem:bit_req_no_cn}
    Set $\lambda' = \frac{c\cdot \|f(t)\|_{\infty}}{\mathrm{OF}-2}$ for some positive constant $c \in [1,\mathrm{OF} - 2]$ and choose $n_0$, $n_1$ such that truncation error is negligible. Let
    \begin{align}
        L_0 = 4\Big\lfloor\frac{N}{2\mathrm{OF}}\Big\rfloor + 4\Big\lfloor\frac{ N}{2\mathrm{OF}}\Big\rfloor\cdot\Big\lfloor\frac{\mathrm{OF}-3}{2}\Big\rfloor.
    \end{align}
    Suppose matrix $\mathbf{V}$ satisfies the RIP of order $L_0$ with a RIP constant denoted $\delta_{L_0}$ and that $\delta_{L_0+1} < \frac{1}{\sqrt{L_0}+1}$. Let $\eta = \|\mathbf{V}^{\mathrm{H}}\mathbf{V}\|_{\infty}$. If $\mathrm{OF} > 3$ and
    \begin{align}\label{eq:bit_req_no_cn}
        b > 3 + \log_2\left\{\frac{3\eta\cdot\frac{\left(\sqrt{1+\delta_{L_0+1}}+1\right)\sqrt{L_0}}{1-(\sqrt{L_0}+1)\delta_{L_0+1}} + 1}{4}\right\},
    \end{align}
   then $c[n]$ can be perfectly recovered using OMP with stopping rule $\|\mathbf{V}^\mathrm{H}\mathbf{r}_t\|_{\infty} \leq \eta\frac{6\lambda'}{2^b-2}$. The $\mathrm{MSE}_{r}$ of the OMP-based reconstruction procedure for the modulo ADC system without the 1-bit folding information signal is
    \begin{align}
        \mathrm{MSE}_{r}
        =& \frac{c^2}{\mathrm{OF}(2^b - 2)^2}\cdot\left(\frac{\|f(t)\|_{\infty}}{\mathrm{OF}-2}\right)^2.
    \end{align}
\end{theorem}
\begin{proof}
    See Appendix \ref{proof:appendix_C}.
\end{proof}

The intuition of the proof is to identify that the smallest non-zero magnitude of $\Delta z[n]$ is $\lambda'$, and that $\epsilon_0 = \eta \frac{6\lambda'}{2^b - 2}$ in \eqref{eq:OMP_req}. The resulting equation enabled us to derive a sufficient condition for the quantization bit.

The main difference between Theorem \ref{theorem:bit_req_no_cn} and Theorem \ref{theorem:quant_noise} is the sufficient condition on the quantization bits to attain the MSE guarantees. The argument inside the logarithm in \eqref{eq:bit_req_no_cn} can be shown to be greater than 1. Thus, the sufficient condition on the quantization bits becomes more stringent in the absence of 1-bit folding information $c[n]$. One major limitation of the bound is that it depends on the RIP constant $\delta_{L_0+1}$. Testing whether a sensing matrix satisfies RIP is an NP-hard problem \cite{Bandeira:2013}. Nevertheless, the sensing matrix $\mathbf{V}_{\mathcal{S}}$ used in our analysis closely resembles a partial DFT matrix. Partial DFT matrices are a class of structured random matrices that are widely studied in compressed sensing and are known to satisfy the RIP with high probability when the number of measurements is sufficiently large relative to the signal sparsity level \cite{Candes:2006}. This theoretical guarantee provides some intuitive justification of the RIP-based analysis in Theorem \ref{theorem:bit_req_no_cn}, even though the RIP constant cannot be evaluated explicitly in practice. Finally, we note that there are other sparse recovery algorithms that can be used to recover $c[n]$. For instance, the LASSO-B$^2$R$^2$ \cite{Shah:2023} uses the iterative soft-thresholding algorithm (ISTA) for sparse recovery. OMP is chosen in this work because of its established theoretical guarantees for perfect support recovery under the bounded noise setting, i.e., Theorem \ref{theorem:amplitude_req}.

\textbf{\textit{Remark}}:
The above analysis provides a sufficient condition on OF and $b$ to ensure that the modulo ADC achieves lower MSE than a conventional ADC. That said, some recent works \cite{UNO, Neuhaus:2021} have explored the use of 1-bit quantizers in both conventional \cite{Neuhaus:2021} and modulo ADCs \cite{UNO} to significantly reduce the ADC's power consumption. Extending our analysis to derive MSE performance guarantees for a modulo ADC with a 1-bit quantizer would therefore be an interesting and valuable direction for future research.

\subsection{Computational Complexity} 
We now derive the computational complexities of our proposed recovery algorithms in Sections \ref{subsection:recovery_method_A} and \ref{subsection:recovery_method_B}. The primary bottleneck of the proposed recovery algorithm for modulo ADC with 1-bit folding information signal is the computation of $\mathbf{V}^{\dagger}_{\mathcal{S}} \in \mathbb{C}^{|\mathcal{S}|\times K}$, where $K$ and $|\mathcal{S}|$ are the number of discrete frequencies in the OOB region and number of folding instances, respectively. For a singular value decomposition (SVD)-based calculation of $\mathbf{V}^{\dagger}_{\mathcal{S}}$, the computational complexity is in $\mathcal{O}\left(K|\mathcal{S}|\cdot\min(K,|\mathcal{S}|)\right)$. Since $|\mathcal{S}| \leq K$ when Lemma \ref{lemma:oversampling_req} holds, the computational complexity of the proposed recovery method in Section \ref{subsection:recovery_method_A} is $\mathcal{O}\left(K|\mathcal{S}|^2\right)$. On the other hand, the proposed recovery algorithm for modulo ADC without the 1-bit folding information signal still requires the same processing steps as the former in addition to the OMP-based support recovery for estimating $c[n]$. The computational complexity of naive implementation of OMP is in $\mathcal{O}\left(NK+K|\mathcal{S}|^2+|\mathcal{S}|^3\right)$, where $N$ is the number of samples within an observation window \cite{Sturm:2012}. Since $\mathcal{O}\left(K|\mathcal{S}|^2\right)$ is in $\mathcal{O}\left(NK+K|\mathcal{S}|^2+|\mathcal{S}|^3\right)$ and $\mathcal{O}(|\mathcal{S}|^3)$ is in $\mathcal{O}(K|\mathcal{S}|^2)$, the computational complexity of the proposed recovery method in Section \ref{subsection:recovery_method_B} is in $\mathcal{O}\left(NK+K|\mathcal{S}|^2\right)$.

The need to estimate the 1-bit folding information $c[n]$ incurs some significant computational cost to the recovery procedure, especially when the modulo ADC is operating at high sampling rates. To show this, consider a BL signal $f(t)$ whose 99\% of its energy is contained in the time window $(-\frac{T_{w}}{2},\frac{T_{w}}{2})$ for some constant $T_{w}$ (in seconds). Suppose further that $f(t) < \lambda'$ outside this time window.  The proposed recovery algorithm needs to process the modulo ADC output samples in this time window so that truncation error will be at most 1\% of the signal energy. For a modulo ADC with sampling rate $f_{s}$, we have $N = \lceil T_{w}f_{s}\rceil$ samples, $K = \lceil(1-\rho)N\rceil$, and $|\mathcal{S}| \leq K$. Thus, the computational complexity of the proposed OMP-based recovery method scales quadratically with the sampling rate. 

The derived computational complexity suggests that the proposed OMP-based recovery could be impractical for high-speed ADCs that need to capture large observation window. Faster sparse recovery algorithms exist such as the iterative soft-thresholding algorithm (ISTA) used in LASSO-$B^2R^2$ \cite{Shah:2023}. We emphasize that the primary focus of our current work is to establish MSE performance guarantees in the presence of quantization noise. Our rationale for selecting OMP is its established theoretical guarantees for perfect support recovery under the bounded noise setting.

\section{Case Study: Simultaneous Acquisition of Weak and Strong Signals}
\label{section:weak_strong}

\begin{figure*}[t!]
    \centering
    \includegraphics[scale = 1]{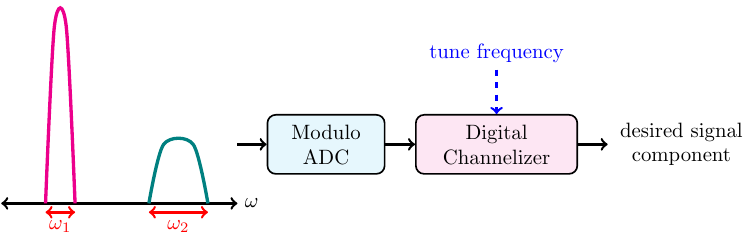}
    \caption{Illustration of simultaneous acquisition of weak and strong signal components using a modulo ADC. The digital channelizer can be tuned to select between the weak and strong components.}
    \label{fig:weak_strong}
\end{figure*}
We now investigate the case of simultaneous acquisition of weak and strong signals that occupy different frequency bands, as shown in Figure \ref{fig:weak_strong}. The modulo ADC block is based on Fig. \ref{fig:modulo_ADC_a} and the digital channelizer implements a digital filter that can be tuned to the desired channel. This case is particularly interesting due to the conflicting requirements to capture the two signals. Setting the dynamic range too high to capture the strong signal can bury the weak signal in quantization noise. Meanwhile, setting the dynamic range too low could distort the strong signal.
This scenario is also of practical importance. For example, hybrid radar fusion systems, such as the one studied in \cite{Chowdary:2024}, involve a similar setup, where strong line-of-sight signals and weak echo signals occupy separate frequency bands.

A recent work on USF \cite{Feuillen:2023} considers the issue of weak and strong components.
However, their mathematical formulation of the problem only assumes a two-tone input signal with different amplitudes. Moreover, performance improvements of modulo ADC over conventional ADCs are only presented via numerical simulations. To the best of our knowledge, a theoretical analysis of the simultaneous acquisition of weak and strong signals using modulo ADC has not been conducted. Our goal in this section is to provide a more general mathematical formulation of the problem and rigorous analysis to justify the advantage of modulo ADC over conventional ADCs in this case.

The bandlimited input signal $f(t)$ is modeled as a two-component signal of the form
\begin{align}
    f(t) = \alpha_1f_1(t) + \alpha_2f_2(t),
    \label{Eq:Two_component_Sig}
\end{align}
where $f_1(t)$ and $f_2(t)$ are bandpass signals. Denote $\mathcal{F}_1$ and $\mathcal{F}_2$ the frequency supports of $f_1(t)$ and $f_2(t)$, respectively. We impose $\mathcal{F}_1\cap\mathcal{F}_2 = \varnothing$ and $\mathcal{F}_1\cup\mathcal{F}_2\in (-\omega_m/2,+\omega_m/2)$, i.e., $f_1(t)$ and $f_2(t)$ have different frequency supports and they both reside inside $(-\omega_m/2,+\omega_m/2)$. The bandwidths of $f_1(t)$ and $f_2(t)$ are $\omega_1$ and $\omega_2$, respectively. Without loss of generality, we set $\|f_1(t)\|_2^2 = 1$ and $\|f_2(t)\|_2^2 = 1$. The simultaneous acquisition of strong and weak signals is modeled by setting $\alpha_1 \gg \alpha_2$. Since $f(t)$ is a bandlimited signal with energy $\alpha_1^2+\alpha_2^2$, we have $\|f(t)\|_{\infty} \leq \sqrt{\frac{\omega_m(\alpha_1^2+\alpha_2^2)}{\pi}}$ by \cite{Papoulis:1967}. 

The modulo ADC system in Fig. \ref{fig:modulo_ADC_a} is adopted in this case study. The reconstruction procedure described in Section \ref{subsection:recovery_method_A} is modified to facilitate recovery of the individual components. Define $\mathrm{BPF}_{i}\{\cdot\}$ to be an ideal digital bandpass filter with bandwidth $\omega_i$ and passband region $\mathcal{F}_i$. After estimating $\hat{z}[n]$, we apply the filter $\mathrm{BPF}_{i}\{\cdot\}$ to $f_{\lambda',\mathrm{q}}[n] - \hat{z}[n]$ to recover the $i$-th signal. Effectively, the recovered signal is
\begin{align*}
    \hat{f}_i[n] = \alpha_if_i[n] +  \mathrm{BPF}_{i}\{z[n] - \hat{z}[n]\} + \mathrm{BPF}_{i}\{\epsilon[n]\}.
\end{align*}
Performance of the modified reconstruction procedure is measured using the normalized MSE (NMSE) incurred when recovering the $i$-th signal, which can be expressed as
\begin{align*}
    \mathrm{NMSE}_{i}=& \frac{\mathbb{E}[(\alpha_if_i[n] - \hat{f}_i[n])^2]}{\alpha_i^2}\\
    =&\frac{1}{\alpha_i^2}\mathbb{E}\Big[(\mathrm{BPF}_{i}\{z[n] - \hat{z}[n]\}+ \mathrm{BPF}_{i}\{\epsilon[n]\})^2\Big].
\end{align*}

At this point, one might consider separating the two signal components in the analog domain using a tunable bandpass filter. This approach enables us to select which signal to capture before the ADC stage. However, tunable filter with high selectivity are costly -- both in terms of hardware resources and design complexity. More precisely, additional analog circuitry is required to vary the center frequency and bandwidth. Tuning elements such as varactors and switched-capacity arrays add complexity, control circuitry, and sensitivity to process, voltage, and temperature (PVT) variations. Furthermore, achieving high selectivity in tunable filters typically requires high-order analog filter designs. However, such high-order filters tend to introduce significant group delay near their cutoff frequencies, leading to signal distortion \cite{harris2021multirate}. In contrast, a tunable digital filter can be implemented by varying the filter coefficients in the registers. Furthermore, achieving a linear phase response is straightforward in digital filters by imposing symmetry on the filter coefficients \cite{harris2021multirate}. 


The following corollary of Theorem \ref{theorem:quant_noise} provides an upper bound for $\mathrm{NMSE}_i$ when $b > 3$ and $\mathrm{OF} > 3$.

\begin{corollary}\label{corollary:weak_strong}
    Suppose $b > 3$, $\mathrm{OF} > 3$, and 1-bit folding information is available to the reconstruction procedure. Then the NMSE incurred when recovering the $i$-th component under the modified reconstruction procedure is
    \begin{align}
        \mathrm{NMSE}_{i}\leq \frac{\omega_i}{\alpha_i^2}\cdot\frac{2}{\mathrm{OF}(2^b - 2)^2}\cdot\left(\frac{1}{\mathrm{OF}-2}\right)^2\cdot\left(\frac{\alpha_1^2+\alpha_2^2}{\pi}\right)
    \end{align}
\end{corollary}
\begin{proof}
    See Appendix \ref{proof:appendix_D}.
\end{proof}
Corollary \ref{corollary:weak_strong} states that $\mathrm{NMSE}_i$ decreases linearly with $\frac{\alpha_i^2}{\omega_i}$. This indicates that, in addition to signal power, bandwidth plays a crucial role in determining recovery performance. Furthermore, if we use conventional ADCs with $b > 3$ and $\mathrm{OF} > 3$, we would get
        \begin{align}\label{eq:weak_strong_no_modulo}
    \mathrm{NMSE}_{i}^{(\mathrm{no-modulo})}
    \leq & \frac{\omega_i}{\alpha_i^2}\cdot\frac{2}{\mathrm{OF}(2^{b}-2)^2}\cdot\frac{(\alpha_1^2+\alpha_2^2)}{\pi}.
    \end{align}
The RHS of \eqref{eq:weak_strong_no_modulo} is strictly larger than the bound given in Corollary \ref{corollary:weak_strong}. In addition, the $\mathrm{NMSE}_i = \mathcal{O}\left(\frac{1}{\mathrm{OF}^3}\right)$ for modulo ADC whereas $\mathrm{NMSE}_i = \mathcal{O}\left(\frac{1}{\mathrm{OF}}\right)$ for conventional ADC. This demonstrates the advantage of using modulo ADC over a conventional ADC for the simultaneous acquisition of weak and strong signals. While the analysis above considered only two signal components with different frequency supports, it can be extended to $K > 2$ signal components as long as they all have disjoint frequency supports. The $\mathrm{NMSE}$ of the $k$-th signal component will still be inversely proportional to $\frac{\alpha_k^2}{\omega_k}$, where $\alpha_k^2$ and $\omega_k$ are the energy and angular frequency of the $k$-th signal component, respectively.



\section{Numerical Results}
\label{section:numerical}

In this section, we perform the following simulations to substantiate the proposed theory. 
The aim is to show that, for the same number of quantization bits used for quantization, the modulo ADC  surpasses the conventional ADC lacking the modulo operator in terms of quantization noise reduction. This is done by analyzing how the mean squared error (MSE), which represents quantization noise power, changes with the oversampling factor (OF).
 
\textbf{Simulation-1:}
In this simulation, we consider the modulo ADC system with side information as depicted in Fig. \ref{fig:modulo_ADC_a}. Specifically, we analyze the variation of MSE, as described in Theorem \ref{theorem:quant_noise} and equation \eqref{eq:no_modulo_error}, as the OF value varies.

Consider the input signal 
\begin{equation}
    f(t) = \sum_{i=1}^{5} A_i \left(\text{sinc}\left(f_m (t-\tau_i)\right)\right)^2,\ \ 0\leq t < T,
\end{equation}
where $\text{sinc}(x) = \frac{\sin(\pi x)}{\pi x}$ for any $x\neq 0$.
This signal entails a maximal frequency component of $f_m=50$ Hz and a temporal duration of $T=1$ second. Random coefficients $A_i$ are drawn uniformly from the interval $[-1,\ 1]$. While time offsets $\tau_i$ are chosen randomly within the interval $\left(\frac{T}{4},\ \frac{3T}{4}\right)$, this is to ensure that the residual signal is negligible outside the interval $(0,\ T)$.
Additionally, we normalize $f(t)$ such that $\|f(t)\|_\infty = 1$. Consequently, the modulo threshold is $\lambda' = \frac{1}{\mathrm{OF}-2}$. 
We consider the number of bits $b$ equal to $4$ and sweep the OF values from $3$ to $15$. The dither signal and uniform quantizer are configured according to these parameters. For each OF value, we compute the MSE between true samples, $f[n]$, and the estimated samples, $\hat{f}[n]$. The experiment is repeated for $20000$ i.i.d. noise realizations and the errors are averaged over all realizations.
\begin{figure}[t!]
    \centering
    \includegraphics[height = 5.8cm, width = 10cm]{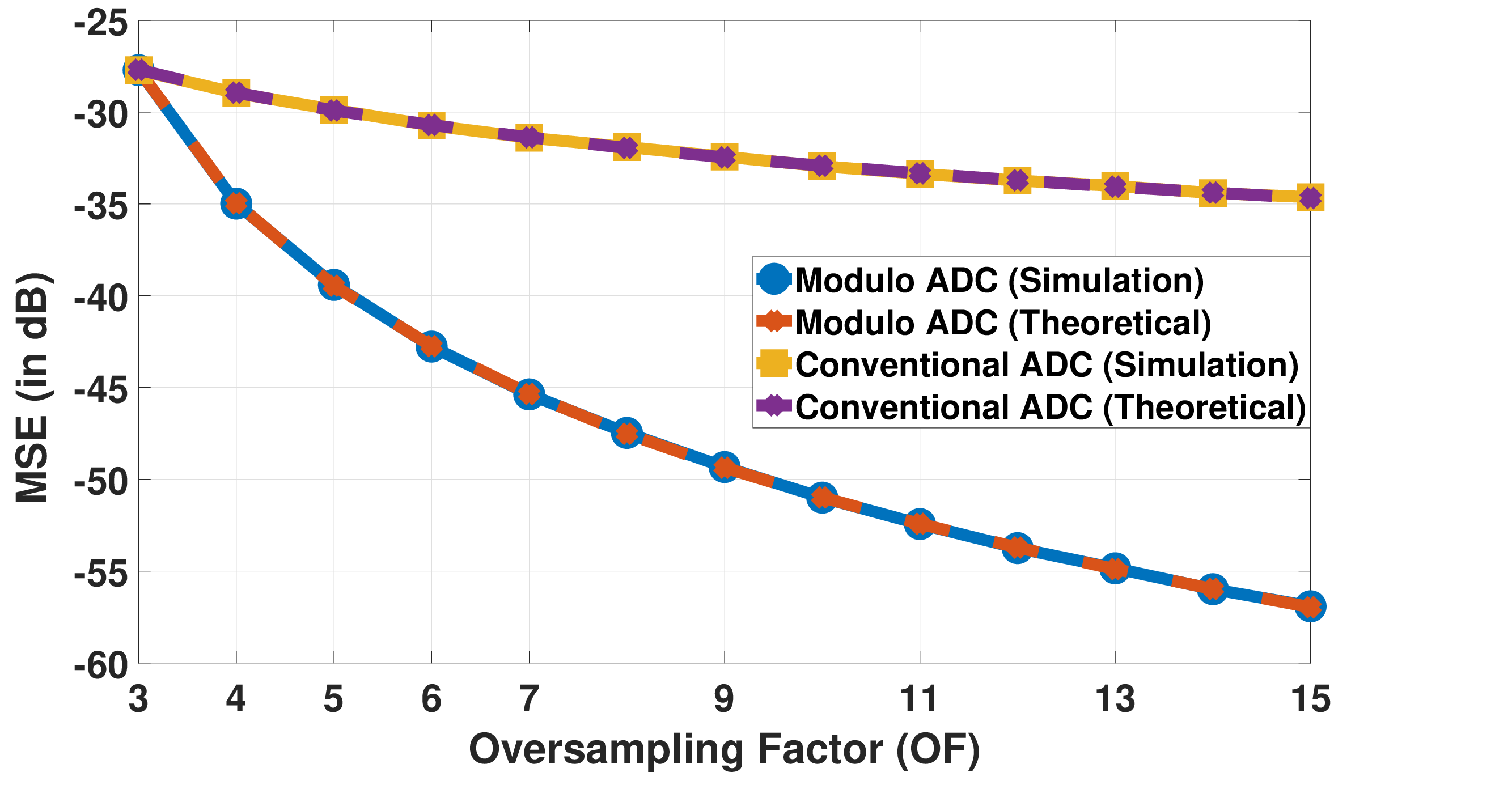}
    \caption{Numerical and theoretical MSE results vs. OF for both modulo ADC with extra 1-bit information and conventional ADC.}
    \label{fig:2}
\end{figure}

Fig. \ref{fig:2} depicts the results obtained with this setting.
It is evident from the figure that the MSE of the $b$-modulo ADC with 1-bit folding information decays much faster than that of the conventional $b$-bit ADC without the modulo operator.
Therefore, from Fig. \ref{fig:2}, we conclude that the quantization noise suppression of modulo ADC is better than that of conventional ADC.
Moreover, it is worth noting that for $\text{OF}=3$, $\lambda'=\|f(t)\|_\infty$, and the MSE values of both modulo ADC and conventional ADC are equal. Fig. \ref{fig:2} also depicts theoretical MSE values obtained using Theorem \ref{theorem:quant_noise} and equation \eqref{eq:no_modulo_error}. 
Note that both theoretical predictions match the simulated results. This justifies the accuracy of our analytical results and further emphasizes the fact that the MSE decays with the order of $\mathcal{O}\left(\frac{1}{\mathrm{OF}^3}\right)$ and $\mathcal{O}\left(\frac{1}{\mathrm{OF}}\right)$ for modulo ADC and conventional ADC, respectively. As a final remark, the use of the extra 1-bit side information in Fig. \ref{fig:modulo_ADC_a} only incurs one additional OR gate to the setup \cite{Shah:2023b}. This penalty is fixed regardless of $b$. From a power efficiency viewpoint, this additional OR gate has minimal impact on the power consumption of modulo ADC.

\begin{figure}[t!]
    \centering
    \includegraphics[height = 5.8cm, width = 10cm]{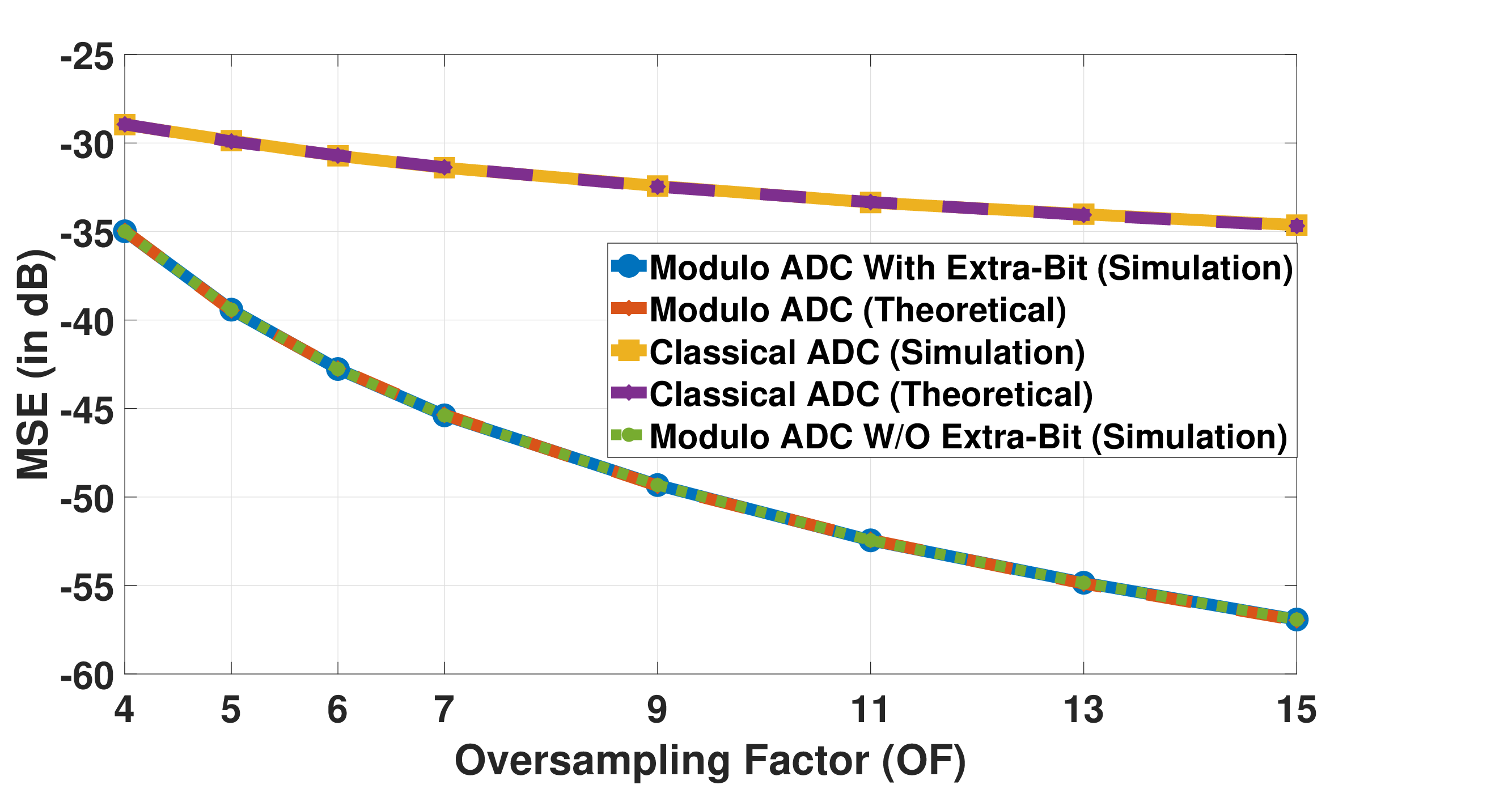}
    \caption{Numerical and theoretical MSE results vs. OF for modulo ADC with and without extra 1-bit side information, and conventional ADC.}
\label{fig:MSE_vs_OF_Without_Extra_Bit}
\end{figure}

\textbf{Simulation-2:}
In this simulation, we examine the modulo ADC system without the additional 1-bit side information. Specifically, we analyze the MSE variation, as detailed in Theorem \ref{theorem:bit_req_no_cn} and equation \eqref{eq:no_modulo_error}, across different OF values. The input signal settings align with those of Simulation-1, featuring $b=4$, OF ranging from $4$ to $15$, an OMP stopping criteria parameter of $\eta =0.8$, and $20000$ i.i.d noise realizations. Fig. \ref{fig:MSE_vs_OF_Without_Extra_Bit} shows the variation of MSE, between true samples and estimated samples, for this setting as OF varies. The figure demonstrates that modulo ADCs with and without extra-bit side information exhibit superior quantization noise suppression compared to the conventional ADC. Furthermore, our theoretical predictions align well with the simulated results; highlighting that the MSE diminishes at rates of $\mathcal{O}\left(\frac{1}{\mathrm{OF}^3}\right)$ and $\mathcal{O}\left(\frac{1}{\mathrm{OF}}\right)$ for the modulo ADC and conventional ADC, respectively.

\textbf{Simulation-3:} 
\begin{figure}[t!]
    \centering
    \includegraphics[height = 5.6cm, width = 8.5cm]{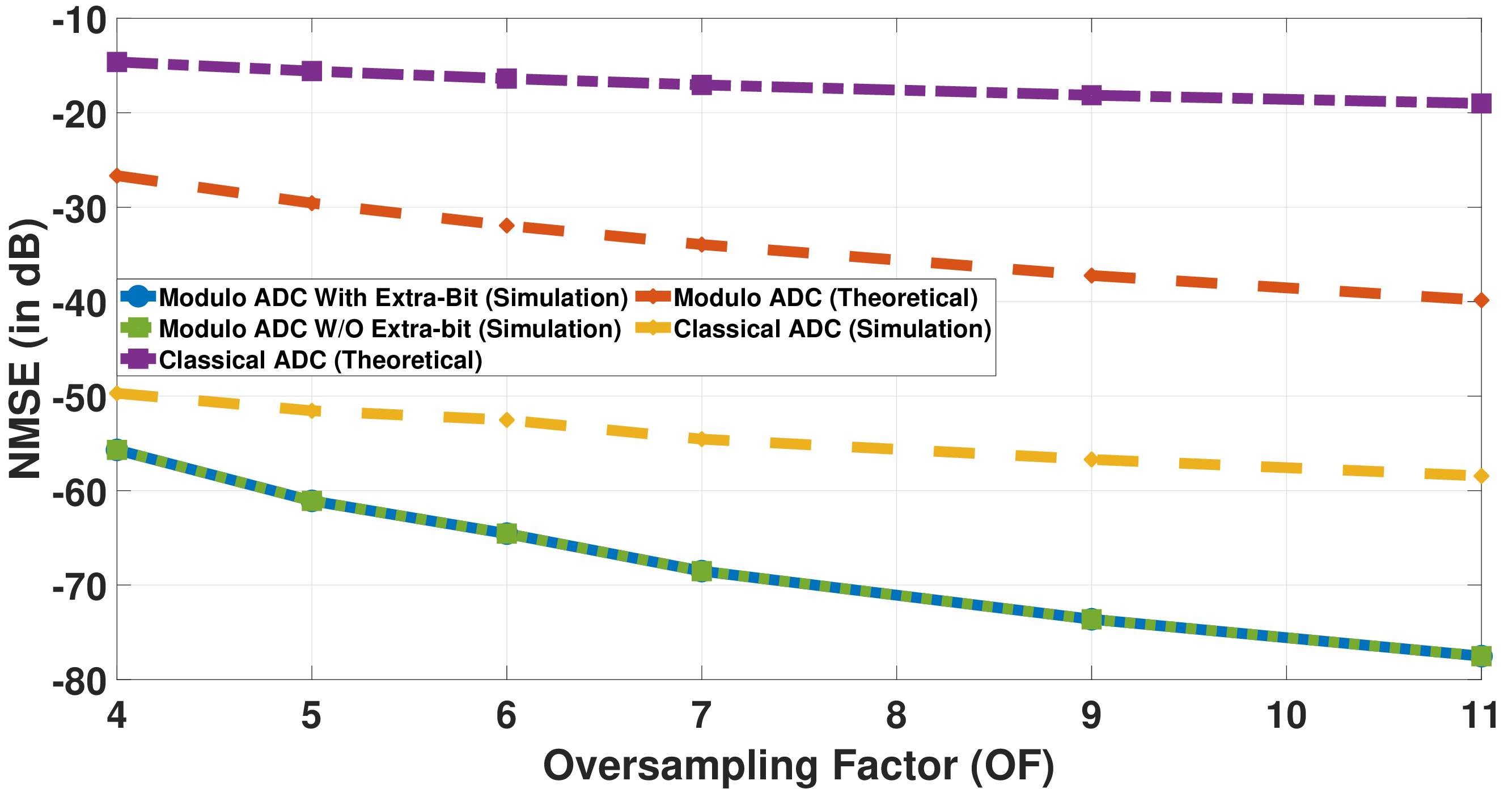}
    \caption{Numerical and theoretical NMSE vs. OF for modulo ADC (with/without extra 1-bit information) and conventional ADC in recovering the strong signal component of a BL signal.}
    \label{fig:MSE_vs_OF_Strong_Sig}
\end{figure}
\begin{figure}[t!]
    \centering
    \includegraphics[height = 5.6cm, width = 8.5cm]{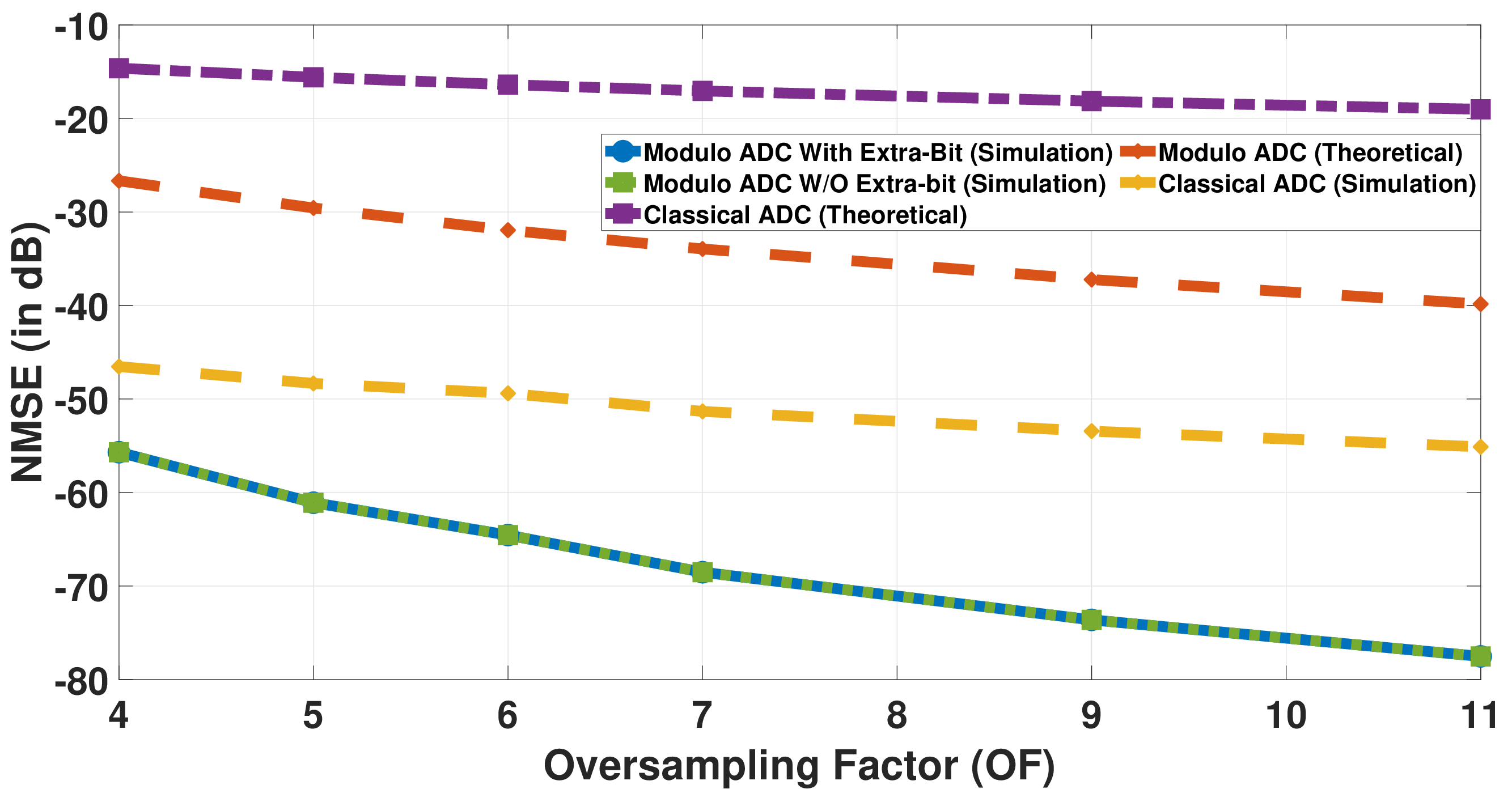}
    \caption{Numerical and theoretical NMSE vs. OF for modulo ADC (with/without extra 1-bit information) and conventional ADC in recovering the weak signal component of a BL signal.}
    \label{fig:MSE_vs_OF_Weak_Sig}
\end{figure}
In this simulation, we analyze modulo ADC systems with and without additional 1-bit information, as well as conventional ADCs, to capture both weak and strong signals occupying distinct frequency bands. Specifically, we validate the upper bound on the NMSE for varying OF values, as stated in Corollary \ref{corollary:weak_strong} and \eqref{eq:weak_strong_no_modulo}, for both strong and weak signal components. The analysis employs the BL signal model described in \eqref{Eq:Two_component_Sig}, with parameters $\alpha_1 = 1$ and $\alpha_2 = 0.25$. The signals $f_1(t)$ and $f_2(t)$ are synthesized using the $\text{sinc}(\cdot)$ function, where the bandwidth of $f_1(t)$ is $[-20 \, \mathrm{Hz}, 20 \, \mathrm{Hz}]$, and the bandwidth of $f_2(t)$ is $[-90 \, \mathrm{Hz}, -50 \, \mathrm{Hz}] \cup [50 \, \mathrm{Hz}, 90 \, \mathrm{Hz}]$.
The simulation parameters are configured as follows: $b = 4$, OF ranging from $4$ to $11$, an OMP stopping criteria parameter of $\eta = 0.7$, and $5000$ i.i.d. noise realizations.
Fig. \ref{fig:MSE_vs_OF_Strong_Sig} and Fig. \ref{fig:MSE_vs_OF_Weak_Sig} illustrate the variation in NMSE across different OF values for recovering the strong and weak signal components, respectively. These results demonstrate that the modulo ADC outperforms the conventional ADC in the simultaneous acquisition of both strong and weak amplitude components that occupy different frequency bands in a BL signal. Furthermore, these simulation results validate the proposed theoretical upper bound on NMSE. Therefore, the NMSE decreases at rates of $\mathcal{O}\left(\frac{1}{\mathrm{OF}^3}\right)$ and $\mathcal{O}\left(\frac{1}{\mathrm{OF}}\right)$ for the modulo ADC and conventional ADC, respectively, during the acquisition of strong and weak signal components in a BL signal.

\begin{figure}[t!]
    \centering
    \includegraphics[height = 5.6cm, width = 10cm]{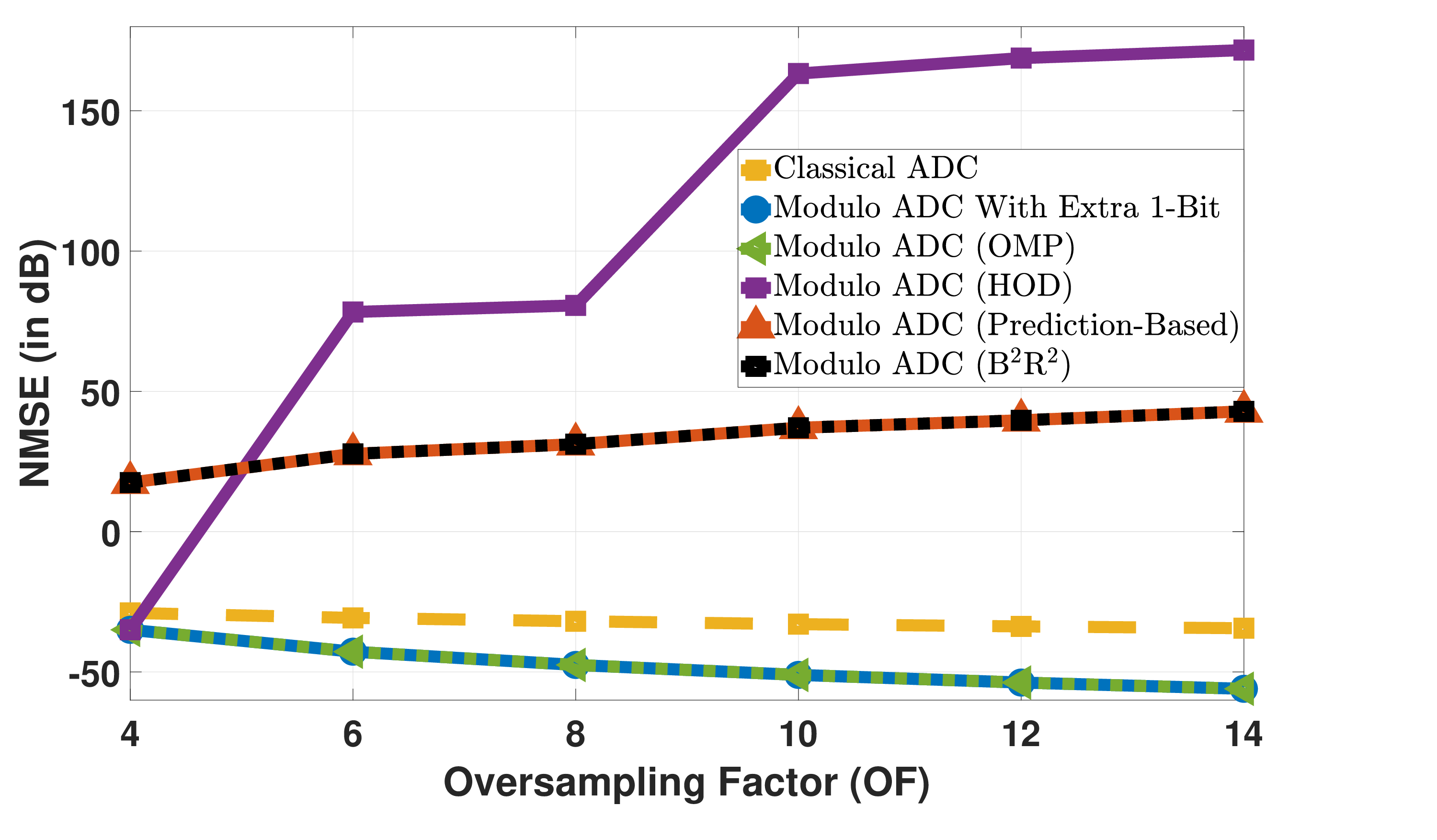}
    \caption{Numerical NMSE vs. OF for modulo and conventional ADCs with $b=4$. Here, the modulo ADC employs various recovery algorithms, including HOD, prediction-based, Fourier-domain-based (B$^2$R$^2$), the proposed OMP-based method, and with extra 1-bit information.}
    \label{fig:MSE_vs_OF_b_4}
\end{figure}

\begin{figure}[t!]
    \centering
    \includegraphics[height = 5.6cm, width = 10cm]{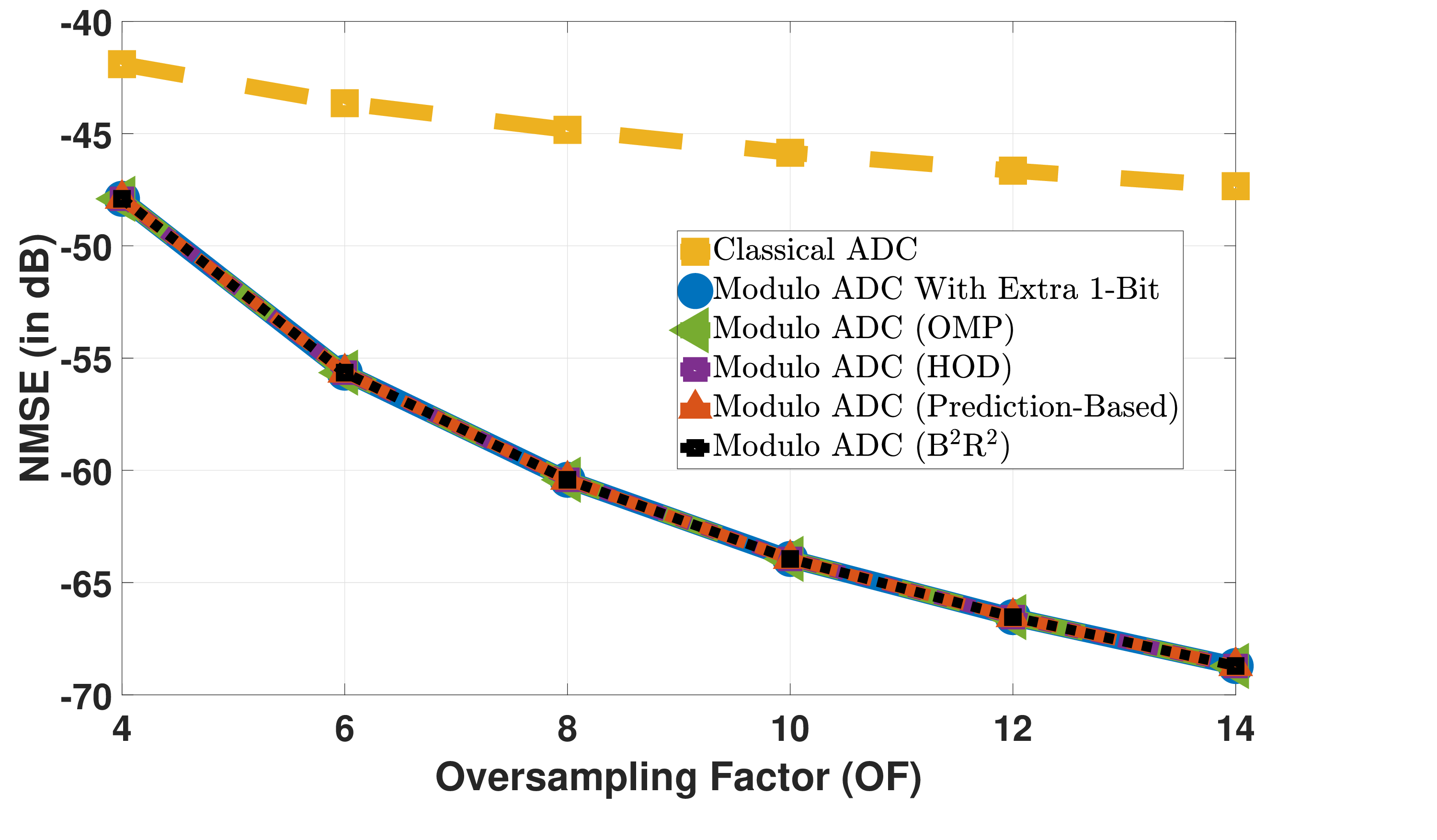}
    \caption{Numerical NMSE vs. OF for modulo and conventional ADCs with $b=6$. Here, the modulo ADC employs various recovery algorithms, including HOD, prediction-based, Fourier-domain-based (B$^2$R$^2$), the proposed OMP-based method, and with extra 1-bit information.}
    \label{fig:MSE_vs_OF_b_6}
\end{figure}

\textbf{Simulation-4:}
In the above simulations, we demonstrated that the modulo ADC outperforms the conventional ADC when using both the proposed OMP-based recovery algorithm and and the algorithm with 1-bit folding information.
As noted in Section \ref{section:intro}, several other recovery algorithms for modulo ADCs have been proposed in the literature, including HOD \cite{US1}, prediction-based \cite{Cheb}, and Fourier-domain-based methods \cite{B2R2, B2R21}. In this simulation, we evaluate and compare the NMSE performance of the modulo ADC using these different recovery algorithms.
We adopt the same simulation setup as in Simulation-1, with $\eta = 0.8$ and 2000 i.i.d. noise realizations. Fig.~\ref{fig:MSE_vs_OF_b_4} plots the NMSE versus OF for $b = 4$, comparing the performance of the conventional ADC and the modulo ADC with various recovery algorithms. Notably, the proposed OMP-based and 1-bit-aided recovery methods outperform all other approaches. The relatively poor performance of HOD can be attributed to its sensitivity to noise.
Furthermore, we observe that when $b = 6$, the performance of the HOD, prediction-based, and B$^2$R$^2$ methods becomes comparable to that of the OMP-based and 1-bit-aided algorithms. These results are presented in Fig.~\ref{fig:MSE_vs_OF_b_6}. Overall, this analysis highlights the superior performance of the proposed OMP-based and 1-bit-aided recovery techniques, particularly under lower quantizer resolutions.


\textbf{Simulation-5:} 
\begin{figure*}[t!]
    \centering
    \includegraphics[height = 6cm, width = 16cm]{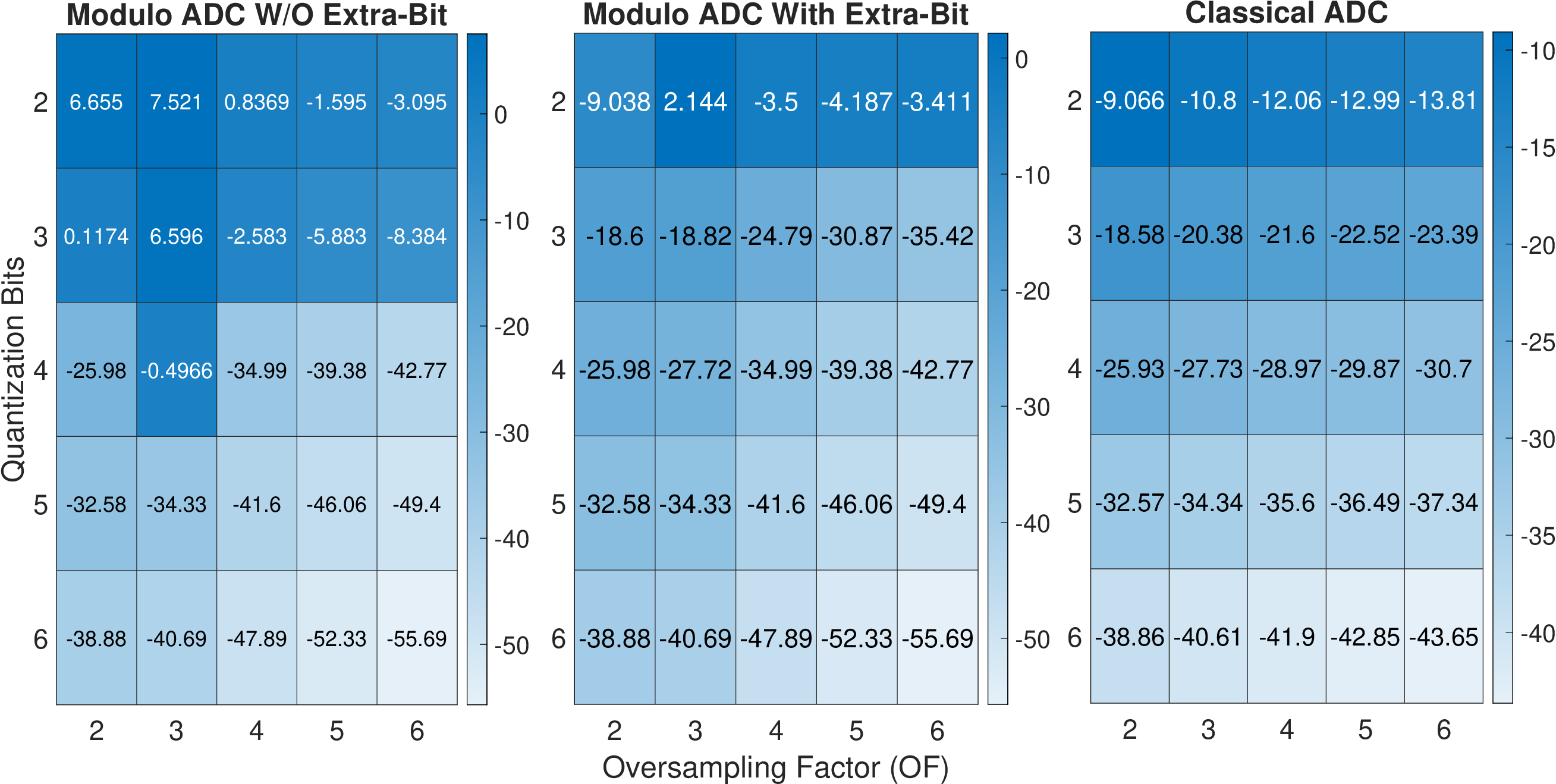}
    \caption{Numerical NMSE results (in dB) for different OF and $b$ values of modulo ADC (with/without extra 1-bit information) and conventional ADC.}
    \label{fig:MSE_vs_OF_vs_b}
\end{figure*}
In the previous simulations, the OF and $b$ were chosen to satisfy the theoretical conditions for accurate recovery.  Here, we examine how the performance of the modulo ADC is affected when these sufficient conditions are not met. Specifically, we compare the NMSE of a classical ADC with that of modulo ADCs, with and without the additional 1-bit information, under various combinations of OF and $b$ that violate the sufficient conditions. The simulation uses the same input signal as in Simulation-1 with $\eta = 0.8$, and averages results over 2000 i.i.d. noise realizations. The NMSE results are shown in Fig.~\ref{fig:MSE_vs_OF_vs_b}, leading to the following observations:
\begin{itemize}
\item For $b = 2$, the classical ADC outperforms the modulo ADC (with and without 1-bit information) for all OF $\geq 2$.
\item For $b = 3$, the classical ADC outperforms the modulo ADC without 1-bit information and performs comparably to the modulo ADC with 1-bit information for all OF $\geq 2$.
\end{itemize}

\begin{figure}[t!]
    \centering
    \includegraphics[height = 5.6cm, width = 8.5cm]{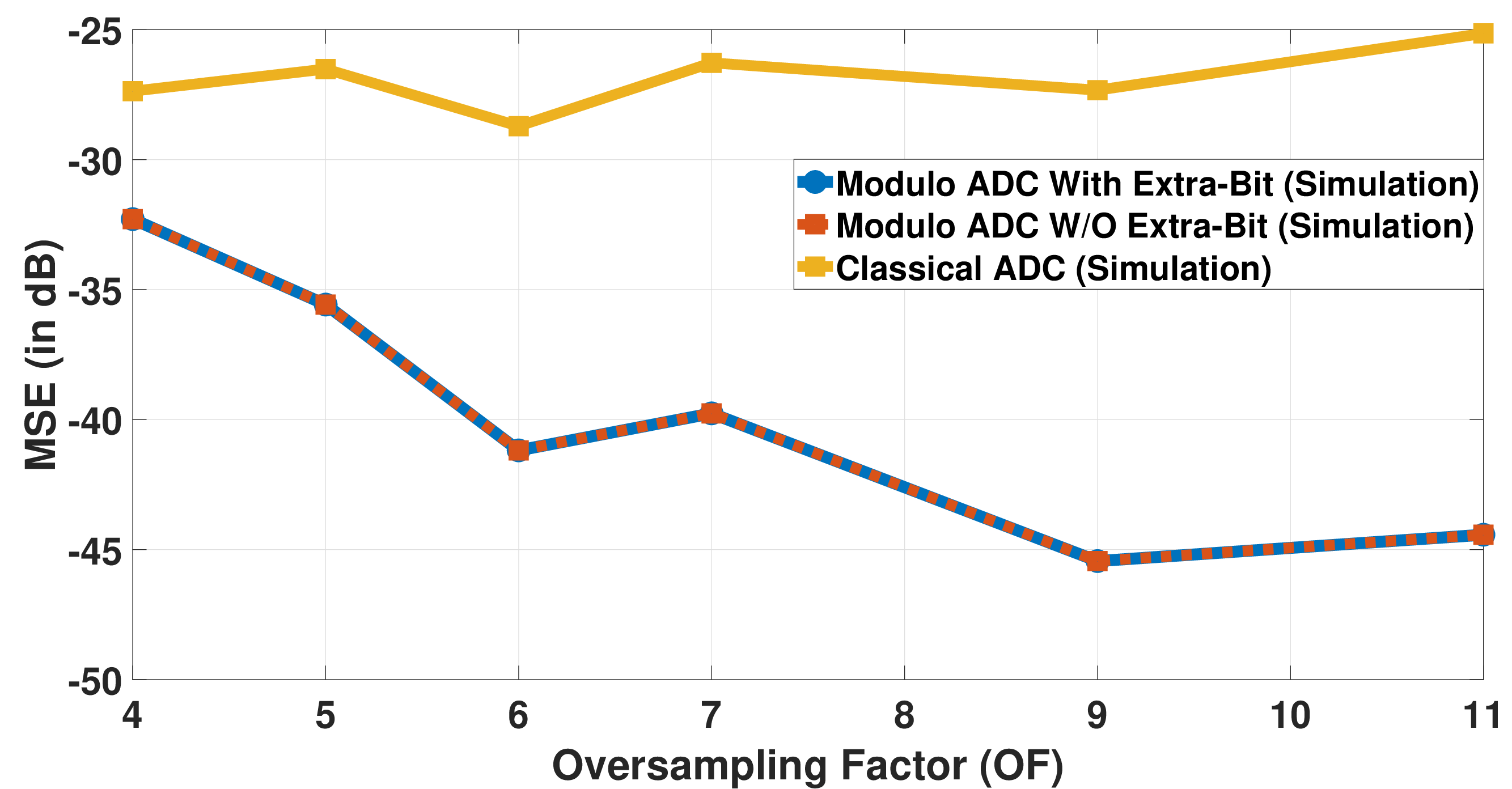}
    \caption{Numerical MSE vs. OF result for modulo ADC with and without extra 1-bit side information, and conventional ADC.}
\label{fig:MSE_vs_OF_True_quantizer}
\end{figure}

\textbf{Simulation-6:}
Above simulations use the dithered quantization framework. Here, we demonstrate that even without a dither signal (i.e., using the conventional quantization framework), the modulo ADC combined with the proposed recovery algorithms outperforms the classical ADC. For this analysis, we consider an input signal with the same settings as Simulation-1: $b=4$, an oversampling factor (OF) ranging from $4$ to $11$, an OMP stopping criterion parameter of $\eta=0.8$, a modulo threshold $\lambda = \frac{1}{\mathrm{OF}-2}$, and $5000$ i.i.d. noise realizations. Fig. \ref{fig:MSE_vs_OF_True_quantizer} illustrates the variation in MSE across different OF values for both modulo and classical ADCs. The results show that the modulo ADC, with or without 1-bit information, consistently outperforms the classical ADC.

\section{Conclusion}
\label{section:conclusion}

In this study, we analyzed the performance of a dithered modulo ADC. We first considered the case where a 1-bit folding information signal is available and proposed a recovery algorithm for unfolding the modulo samples. Our analysis showed that the conditions $\mathrm{OF} > 3$ and $b > 3$ are sufficient for achieving better quantization noise suppression compared to a conventional ADC with the same amplitude quantization bit budget. Additionally, we demonstrated that the MSE of the modulo ADC decreases more rapidly with increasing $\mathrm{OF}$ than that of a conventional ADC. Recognizing that folding information is often unavailable in practical implementations, we also developed and analyzed an orthogonal matching pursuit (OMP)-based reconstruction algorithm that does not require an additional folding information signal. The proposed OMP-based recovery method achieves the same MSE performance as that of the original unfolding scheme but the lack of 1-bit folding information penalizes the sufficient condition for $b$. More precisely, the sufficient condition for the oversampling factor is still $\mathrm{OF} > 3$ but the amplitude quantization bit budget becomes $b > 3 + \log_2(\zeta)$, where $\zeta$ is the argument of the $\log_2(\cdot)$ operator in \eqref{eq:bit_req_no_cn}. We also analyzed the case of simultaneous acquisition of weak and strong signals occupying different frequency bands and demonstrated the superior performance of modulo ADCs over conventional ADCs. Numerical results are provided to substantiate the performance guarantees derived in this work.

\begin{appendices}
\section{Proof of Lemma \ref{lemma:oversampling_req}}\label{proof:appendix_A}

A necessary condition for $\mathbf{V}_{\mathcal{S}}$ to be full column rank is $|\mathcal{S}|\leq K$. Since the columns of $\mathbf{V}_{\mathcal{S}}$ are from the columns of the Fourier basis, they are linearly independent. Thus, the condition $|\mathcal{S}|\leq K$ is also sufficient.

    We first provide an upper bound for $|\mathcal{S}|$. Since $n_0$ and $n_1$ are chosen such that truncation error is negligible, we can use the results of \cite{Shah:2023} regarding the sparsity of the first-order difference of the modulo residue. Using \cite[Equation 7]{Shah:2023}, we have
    \begin{align}\label{eq:S_upperbound}
        |\mathcal{S}| \leq& 4\Big\lfloor\frac{\rho N}{2}\Big\rfloor + 4\Big\lfloor\frac{\rho N}{2}\Big\rfloor\cdot\Big\lfloor\frac{\|f(t)\|_{\infty}-\lambda'}{2\lambda'}\Big\rfloor\nonumber\\
        \leq & 2\rho N + 2\rho N\cdot\left(\frac{\|f(t)\|_{\infty}-\lambda'}{2\lambda'}\right),
    \end{align}
    where the second inequality comes from the trivial upper bound of the floor function, i.e., $\lfloor x\rfloor \leq x$. Meanwhile, $K = N - 2\lfloor\frac{\rho N}{2}\rfloor \geq (1-\rho)N$. Hence, for $\mathbf{V}_{\mathcal{S}}$ to be a full column rank matrix, it suffices to show that
    \begin{align*}
           2\rho N + 2\rho N\cdot\left(\frac{\|f(t)\|_{\infty}-\lambda'}{2\lambda'}\right) \leq (1-\rho)N.
    \end{align*}
    After some algebraic manipulation, we get
    \begin{align}\label{eq:lambda_requirement}
        \lambda' \geq \frac{\rho}{1-2\rho}\|f(t)\|_{\infty} \quad\left(\text{or } \lambda' \geq \frac{\|f(t)\|_{\infty}}{\mathrm{OF}-2}  \right).
    \end{align}
    Since $\lambda' < \|f(t)\|_{\infty}$, there exists a $\lambda'$ that satisfies the above inequality when $\rho < \frac{1}{3}$, i.e., $\mathrm{OF} > 3$.

\section{Proof of Theorem \ref{theorem:quant_noise}}\label{proof:appendix_B}

We first bound the $\ell_\infty$-norm of the estimate of the first-order difference of the modulo residue:
\begingroup
\allowdisplaybreaks
    \begin{align*}
        \|{\Delta}\mathbf{\hat{z}} - {\Delta}\mathbf{z}\|_{\infty} =& \|\mathbf{V}_{\mathcal{S}}^{\dagger}\mathbf{y} - {\Delta}\mathbf{z}_{\mathcal{S}}\|_{\infty}\\
        =& \|\mathbf{V}_{\mathcal{S}}^{\dagger}\mathbf{V}_{\mathcal{S}}\left({\Delta}\mathbf{z}_{\mathcal{S}} + {\Delta}\boldsymbol{\epsilon}_{\mathcal{S}}\right) - {\Delta}\mathbf{z}_{\mathcal{S}}\|_{\infty}\\
        =&\|{\Delta}\boldsymbol{\epsilon}_{\mathcal{S}}\|_{\infty}\\
        \leq& \frac{6\lambda}{2^b} \\
        =& \frac{6\lambda'}{2^b-2},
    \end{align*}
    \endgroup
    where ${\Delta}\boldsymbol{\epsilon}_{\mathcal{S}}$ contains the values of $\Delta\epsilon[n]$ in the indices contained in $\mathcal{S}$. The third line follows because $\mathbf{V}_{\mathcal{S}}$ is the full column rank according to Lemma \ref{lemma:oversampling_req}. The fourth line holds because $\epsilon[n]$ induced by the triangle dither $d[n]$ has amplitude support $\left(-\frac{3\lambda}{2^b},\;+\frac{3\lambda}{2^b}\right)$. Hence, $|\Delta\epsilon[n]| \leq \frac{6\lambda}{2^b}$ for all $n$.

    To ensure that we perfectly recover the modulo residue after the rounding operation, we must have
    \begin{align}
        \frac{6\lambda'}{2^b-2} < \lambda'\quad \Longrightarrow\quad b > 3,
    \end{align}
    which is an explicit assumption of the theorem. In this setting, $z[n] - \hat{z}[n] = 0$. Consequently, from \eqref{eq:recovered_sig}, the MSE becomes
    \begin{align}
        \mathrm{MSE} =& \mathbb{E}\left[(\mathrm{LPF}\{\epsilon[n]\})^2\right]
        \nonumber\\
        =& \rho\frac{\lambda^2}{2^{2b}} \nonumber\\
        =& \frac{\lambda'^2}{\mathrm{OF}(2^b - 2)^2}.
    \end{align}
    Since MSE grows quadratically with $\lambda'$, we should set the value of $\lambda'$ as low as possible. Noting that $\lambda'$ should satisfy \eqref{eq:lambda_requirement}, we obtain the desired result by setting $\lambda' = \frac{c\cdot \|f(t)\|_{\infty}}{\mathrm{OF}-2}$ for some positive constant $1 \leq c \leq \mathrm{OF} - 2$.

\section{Proof of Theorem \ref{theorem:bit_req_no_cn}}\label{proof:appendix_C}

Recall that finding the unknown support $\mathcal{S}$ of $\Delta z[n]$ is the same as finding $c[n]$. We focus on the former. Since the nonzero elements of the modulo residue have amplitudes $(2\mathbb{Z}+1)\lambda'$, we get
    \begin{align*}
        \min_{n\in\mathcal{S}} |\Delta z[n]| = \lambda'. 
    \end{align*}
    In addition, 
    \begin{align*}
        \|\mathbf{V}^{\mathrm{H}}\mathbf{V}\Delta\boldsymbol{\epsilon}\|_{\infty} \leq & \|\mathbf{V}^{\mathrm{H}}\mathbf{V}\|_{\infty}\|\Delta\boldsymbol{\epsilon}\|_{\infty}\\
        \leq & \eta\frac{6\lambda'}{2^b-2}
    \end{align*}
    By Theorem \ref{theorem:amplitude_req}, exact support recovery is guaranteed if
    \begin{align*}
       & \lambda' > \frac{\left(\sqrt{1+\delta_{L_0+1}}+1\right)\sqrt{L_0}}{1-(\sqrt{L_0}+1)\delta_{L_0+1}}\left(\eta\frac{6\lambda'}{2^b-2}\right)\\
        \Rightarrow & b > \log_2\left\{6\eta\cdot\frac{\left(\sqrt{1+\delta_{L_0+1}}+1\right)\sqrt{L_0}}{1-(\sqrt{L_0}+1)\delta_{L_0+1}} + 2\right\}\\
        &\;\; = 3 + \log_2\left\{\frac{3\eta\cdot\frac{\left(\sqrt{1+\delta_{L_0+1}}+1\right)\sqrt{L_0}}{1-(\sqrt{L_0}+1)\delta_{L_0+1}} + 1}{4}\right\}.
    \end{align*}
    With support $\mathcal{S}$ perfectly recovered, the signal recovery block effectively has the 1-bit folding information signal $c[n]$. The proof is completed by applying the theoretical guarantees in Theorem \ref{theorem:quant_noise}.

\section{Proof of Corollary \ref{corollary:weak_strong}}\label{proof:appendix_D}

From Theorem \ref{theorem:quant_noise}, $z[n] - \hat{z}[n] = 0$ if $b > 3$ and $\mathrm{OF} > 3$. The normalized MSE of recovering the $i$-th component becomes
    \begin{align}
        \frac{\mathbb{E}[(\alpha_if_i[n] - \hat{f}_i[n])^2]}{\mathbb{E}[(\alpha_if_i[n])^2]} =& \frac{1}{\alpha_i^2}\mathbb{E}[(\mathrm{BPF}_{i}\{\epsilon[n]\})^2]\nonumber\\
        = & \frac{\rho}{\alpha_i^2}\cdot \frac{\omega_i}{\omega_m}\cdot\frac{2\lambda'^2}{(2^{b}-2)^2}.
    \end{align}
    Setting $\lambda'$ to be the RHS of \eqref{eq:lambda_requirement} and replacing $\|f(t)\|_{\infty}$ by its upper bound $\sqrt{\frac{\omega_m(\alpha_1^2+\alpha_2^2)}{\pi}}$ complete the proof.

\end{appendices}

\renewcommand*{\bibfont}{\footnotesize}
\begingroup
\footnotesize  
\printbibliography
\endgroup

\vspace{-1cm}
\begin{IEEEbiography}[{\includegraphics[width=1in,height=1.25in,clip,keepaspectratio]{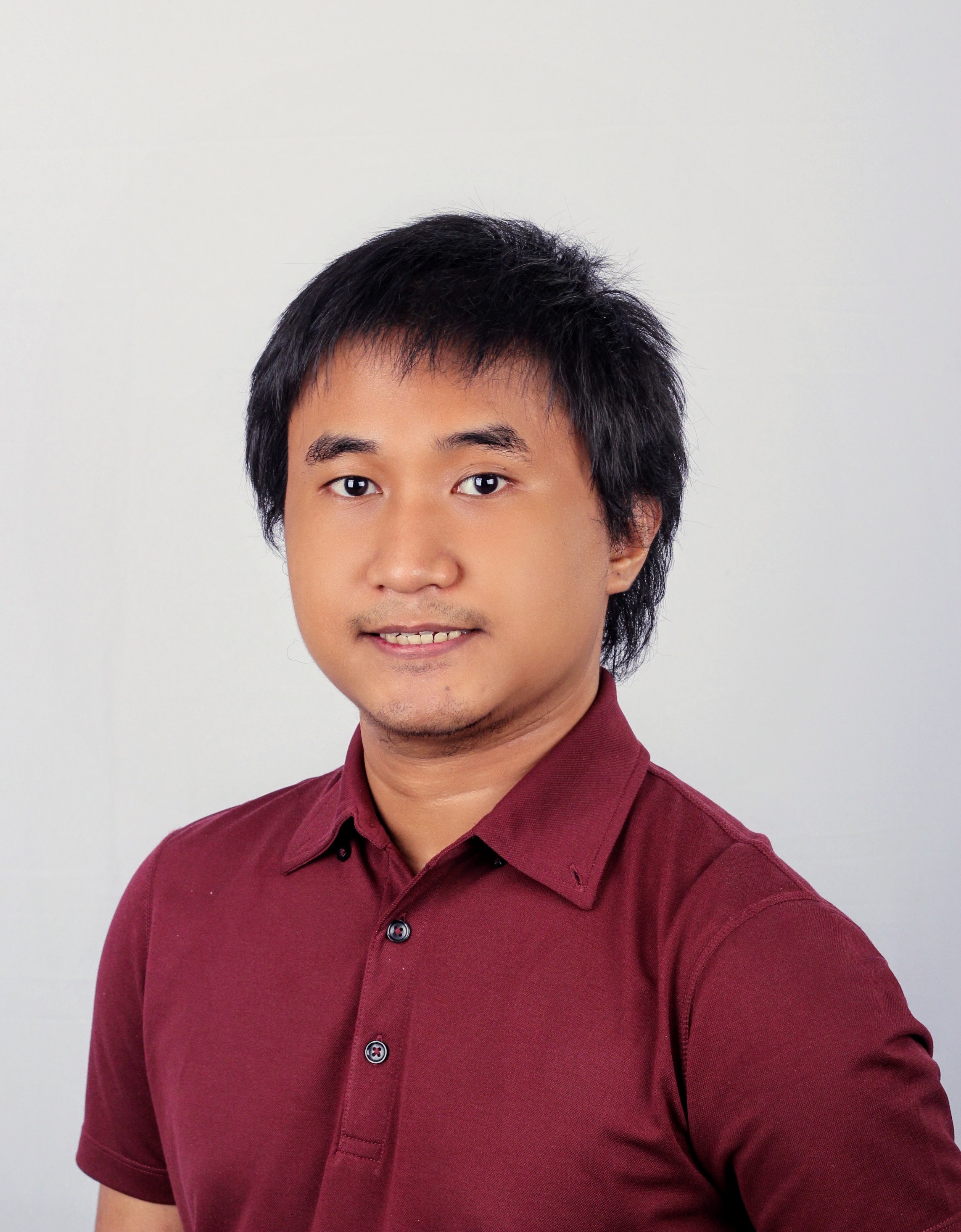}}]{Neil Irwin Bernardo} received the Bachelor of Science degree in Electronics and Communications Engineering and the Master of Science in Electrical Engineering from the University of the Philippines Diliman in 2014 and 2016, respectively. He then received the Ph.D. degree in Engineering from the University of Melbourne, Australia. He is currently an Associate Professor at the Electrical and Electronics Engineering Institute of the University of the Philippines Diliman and is the head of the UP Wireless Communications Engineering Laboratory. His research interests include wireless communications, signal processing, and information theory.
\end{IEEEbiography}

\begin{IEEEbiography}[{\includegraphics[width=1in,height=1.3in,clip,keepaspectratio]{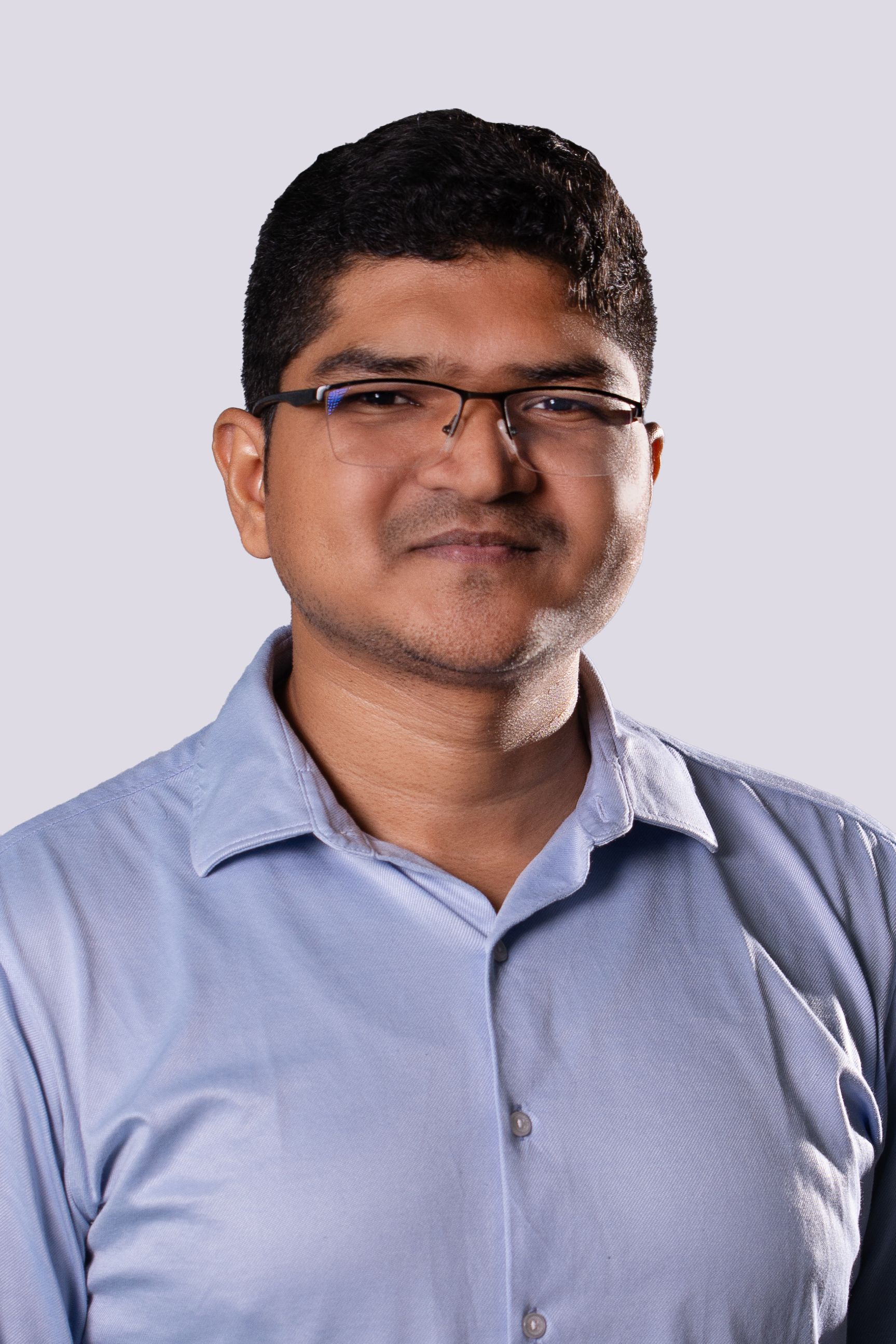}}]{Shaik Basheeruddin Shah} (Graduate student member, IEEE) is a postdoctoral fellow in the Department of Electrical Engineering at Khalifa University, Abu Dhabi, UAE. He earned his Master’s degree in Computational Engineering from Rajiv Gandhi University of Knowledge Technologies, India, in 2015, and his Ph.D. in Electrical Engineering from Shiv Nadar University, India, in 2022. He was awarded the Pratibha Award for excellence in his Master’s studies. From 2022 to 2024, he was a postdoctoral researcher at the Weizmann Institute of Science, where he received the Faculty Dean Fellowship. His research interests include modulation schemes, signal representations, sampling theory, biomedical signal analysis, compressive sensing, and model-based learning.
\end{IEEEbiography}
\begin{IEEEbiography}[{\includegraphics[width=1in,height=1.2in]{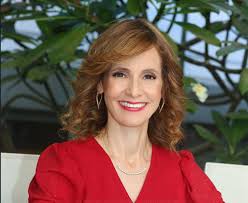}}] {Yonina C. Eldar} (Fellow, IEEE) received the B.Sc. degree in physics and the B.Sc. degree in electrical engineering from Tel Aviv University, Tel-Aviv, Israel, 1995 and 1996, respectively, and the Ph.D. degree in electrical engineering and computer science from the Massachusetts Institute of Technology (MIT), Cambridge, MA, USA, in 2002. She was a Visiting Professor at Stanford University. She is currently a Professor with the Department of Mathematics and Computer Science, Weizmann Institute of Science, Rehovot, Israel, where she heads the Center for Biomedical Engineering and Signal Processing and holds the Dorothy and Patrick Gorman Professorial Chair. She is also a Visiting Professor at MIT, a Visiting Scientist at the Broad Institute, and an Adjunct Professor at Duke University. She is a member of the Israel Academy of
Sciences and Humanities and a EURASIP Fellow. She has received many awards for excellence in research and teaching, including the IEEE Signal Processing Society Technical Achievement Award (2013), the IEEE/AESS Fred Nathanson Memorial Radar Award (2014), and the IEEE Kiyo Tomiyasu Award (2016). She was a Horev Fellow of the Leaders in Science and Technology Program at the Technion and an Alon Fellow. She received the Michael Bruno Memorial Award from the Rothschild Foundation, the Weizmann Prize for Exact Sciences, the Wolf Foundation Krill Prize for Excellence in Scientific Research, the Henry Taub Prize for Excellence in
Research (twice), the Hershel Rich Innovation Award (three times), and the Award for Women with Distinguished Contributions. She received several best paper awards and best demo awards together with her research students and colleagues, was selected as one of the 50 most influential women in Israel, and was a member of the Israel Committee for Higher Education. She is the
Editor-in-Chief of Foundations and Trends in Signal Processing, a member of several IEEE Technical Committees and Award Committees, and heads the Committee for Promoting Gender Fairness in Higher Education Institutions in Israel.
\end{IEEEbiography}







\end{document}